\documentclass[%
reprint,
showpacs,preprintnumbers,
showkeys,
amsmath,amssymb,
aps,
prl,
floatfix,
]{revtex4-1}

\usepackage[english]{babel}
\usepackage[utf8]{inputenc}
\usepackage{graphicx}
\usepackage{hyperref}
\usepackage{xcolor}
\usepackage{amsmath}
\usepackage{amsfonts}
\usepackage{amssymb}
\usepackage{dsfont}
\usepackage{pdfpages}

\definecolor{refColor}{HTML}{EA00F2}
\definecolor{figColor}{HTML}{008DF2}
\definecolor{urlColor}{HTML}{00AEF2}

\hypersetup{
    unicode=false,                 
    pdftoolbar=true,               
    pdfmenubar=true,               
    pdffitwindow=false,            
    pdfstartview={FitH},           
    pdftitle={},                   
    pdfauthor={Nicolai Lang},      
    pdfsubject={Quantum physics},  
    pdfcreator={Nicolai Lang},     
    pdfproducer={Nicolai Lang},    
    pdfkeywords={},                
    pdfnewwindow=true,             
    colorlinks=true,               
    linkcolor=figColor,            
    citecolor=refColor,            
    filecolor=magenta,             
    urlcolor=urlColor              
}

\newcommand{\Bra}[1]{\left<#1\right|}
\newcommand{\Ket}[1]{\left|#1\right>}

\renewcommand{\vec}[1]{\boldsymbol{#1}}

\newcommand{\tr}[1]{\operatorname{Tr}\left[#1\right]}

\newcommand{\Tr}[2]{\operatorname{Tr}_{#2}\left[#1\right]}

\begin{document}


\title{Topological states in a microscopic model of interacting fermions}

\author{Nicolai Lang}
\email{nicolai@itp3.uni-stuttgart.de}
\author{Hans Peter B\"uchler}
\affiliation{Institute for Theoretical Physics III, 
  University of Stuttgart, 70550 Stuttgart, Germany}

\date{\today}


\begin{abstract}
  We present a microscopic model of interacting fermions where the ground state degeneracy is topologically protected. 
  The model is based on a double-wire setup with local interactions in a particle number conserving setting.  
  A compelling property of this model is the exact solvability for its ground states and low energy excitations. 
  We demonstrate the appearance of topologically protected edge states and derive their braiding properties on a microscopic level.  
  We find the non-abelian statistics of Ising anyons, which can be interpreted as Majorana-like edge states.
\end{abstract}

\pacs{03.65.Vf,03.75.Ss,03.67.Lx,74.25.-q}


\maketitle


Topologically protected ground state degeneracies in many-body quantum systems, and the closely related 
(non-abelian) anyonic statistics, are of special interest from a theoretical point of view \cite{Arovas1984,Levin2003}, 
and have been recognized as promising concepts for scalable fault-tolerant quantum computation \cite{1_kitaev_FTQCBA_A,Nayak2008}.
A well understood class are topological states with Majorana zero-energy edge modes appearing within mean-field descriptions 
of topological superconductors \cite{Beenakker2014a}. These free fermion theories have been classified exhaustively \cite{Kitaev2009a,Ryu2010a}, 
and the properties of the Majorana zero modes at boundaries \cite{Kitaev2007} and in vortices \cite{Ivanov2001} 
have been characterized. In contrast, interacting and gapless phases are less well understood \cite{Fidkowski2011,Schuch2011,Chen2011,Wen2014}, 
and to which extent existence and non-abelian properties of edge states carry over to interacting theories 
is an interesting question lacking conclusive answers \cite{Bonderson2013}.

The understanding of topological states is driven by exactly solvable microscopic models; the paradigmatic one
for the existence of topologically protected Majorana edge modes is the one-dimensional Majorana chain \cite{Kitaev2007}. 
It has inspired a variety of proposals for its experimental realization in condensed matter systems \cite{Lutchyn2010,Oreg2010,Alicea2012,Beenakker2014a},
and signatures consistent with Majorana modes have been experimentally observed \cite{Mourik2012,Das2012,Perge2014,Xu2015}. 
Nevertheless, these models require large reservoirs to justify their mean-field description, 
whereas very little is known about the fate of Majorana zero-energy edge modes in intrinsically interacting and particle conserving settings.
Previous attempts for number-conserving theories featuring Majorana-like edge states relied either on 
bosonization~\cite{Cheng2011,Fidkowski2011a,Sau2011,Ruhman,Keselman2015} or on numerical methods (DMRG)~\cite{Kraus2013a}, 
while the only exactly solvable models require unphysical long-range interactions~\cite{Ortiz2014}.

In this letter, we present a microscopic, number-conserving theory with \textit{local} interactions 
that features non-abelian edge states at the boundaries. The theory allows for an exact derivation of its many-body 
ground state as well as its low energy excitations, and thereby provides a viable playground for analyzing its characteristic properties. 
We find that the ground state is characterized by a condensate of $p$-wave pairs with a topological degeneracy. 
The Green's function exhibits a revival at the edges, indicating the appearance of edge states. 
Remarkably, the system can be extended to arbitrary wire networks, which allows us to derive the 
non-abelian braiding statistics of the edge states on a microscopic level.


\begin{figure}[t]
  \centering
  \includegraphics[width=0.75\linewidth]{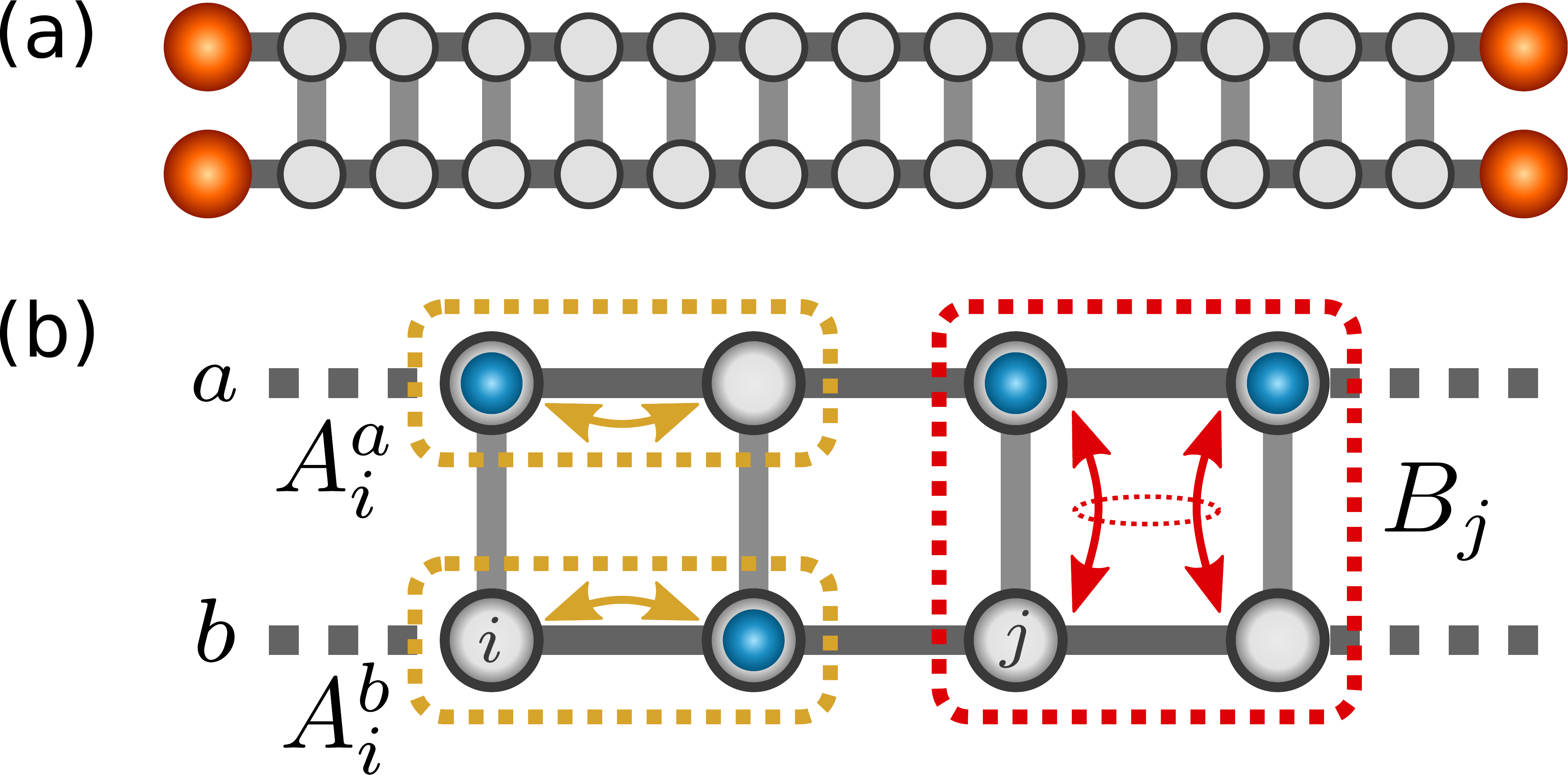}
  \caption{
    \textit{Setup.}
    (a) We consider a double chain (two-leg ladder) of spinless fermions with upper/lower chain denoted as $a$/$b$.
    (b) The number-conserving Hamiltonian is given by intra-chain terms $A_i^x$ ($x=a,b$) and inter-chain couplings $B_j$.
  }
  \label{fig:1}
\end{figure}

We consider a double chain (two-leg ladder) of spinless fermions with $L$ lattice sites. The fermionic creation operators at site $i$ 
are described by $a^{\dag}_{i}$ (upper chain) and $b^{\dag}_{i}$ (lower chain), see Fig.~\ref{fig:1}.
The many-body Hamiltonian  $H= H^a+H^b+H^{ab}$ describing the interacting fermion theory combines intra-chain contributions 
$H^x$ ($x=a,b$) as well as interactions $H^{ab}$ between the two chains. The intra-chain Hamiltonian takes the form
\begin{equation}
  \label{eq:1}
  \textstyle H^x=\sum_{i=1}^{L-1}\, A^x_i\left(\mathds{1}+A^x_i\right) 
\end{equation}
with the single-particle hopping terms
\begin{equation}
  \label{eq:2}
  \textstyle A^a_i=a_ia_{i+1}^\dag +a_{i+1}a_i^\dag\,,\qquad A^b_i=b_i b_{i+1}^\dag +b_{i+1}b_i^\dag\,.
\end{equation}
Consequently, it combines single particle hopping with a nearest-neighbor attraction $n_i^x+n_{i+1}^x-2n_{i}^xn_{i+1}^x$.
The inter-chain interaction $H^{ab}$ takes a similar form
\begin{equation}
  \label{eq:3}
  \textstyle H^{ab}=\sum_{i=1}^{L-1}\, B_i\left(\mathds{1}+B_i\right) 
\end{equation}
with the pair-hopping between the two chains
\begin{equation}
  \label{eq:4}
  \textstyle B_i=a_i^\dag a_{i+1}^\dag b_{i}b_{i+1}+b_{i}^\dag b_{i+1}^\dag a_i a_{i+1}\,.
\end{equation}
It is important to stress that the Hamiltonian $H$ conserves the total number of particles $N$,
which defines the only free parameter of the theory and is conveniently expressed as the filling $\rho=N/2L$.
$H$ features two additional, relevant symmetries, namely (i) the subchain-parity 
$P_x\equiv(-1)^{\sum_i\,x_i^\dag x_i}$ ($x=a,b$), and (ii) time-reversal symmetry $\mathcal{T}\equiv K$ 
represented by complex conjugation $K$ and $\mathcal{T}x_i^{(\dag)}\mathcal{T}^{-1}\equiv x_i^{(\dag)}$.


{\it Ground states} --- 
In order to derive the ground states analytically, the observation that Hamiltonian $H$ 
is the sum of local projectors and therefore a locally positive operator is crucial.
Then we exploit the fact that any zero-energy ground state must be annihilated by all local terms in (\ref{eq:1}) and (\ref{eq:3}) simultaneously.
That is, \textit{if} we find a state with zero energy which is annihilated by all local terms, we can be sure that it is a ground state.
This yields a viable method to construct them from scratch --- provided zero-energy ground states exist. 

\begin{figure}[t]
  \centering
  \includegraphics[width=1.0\linewidth]{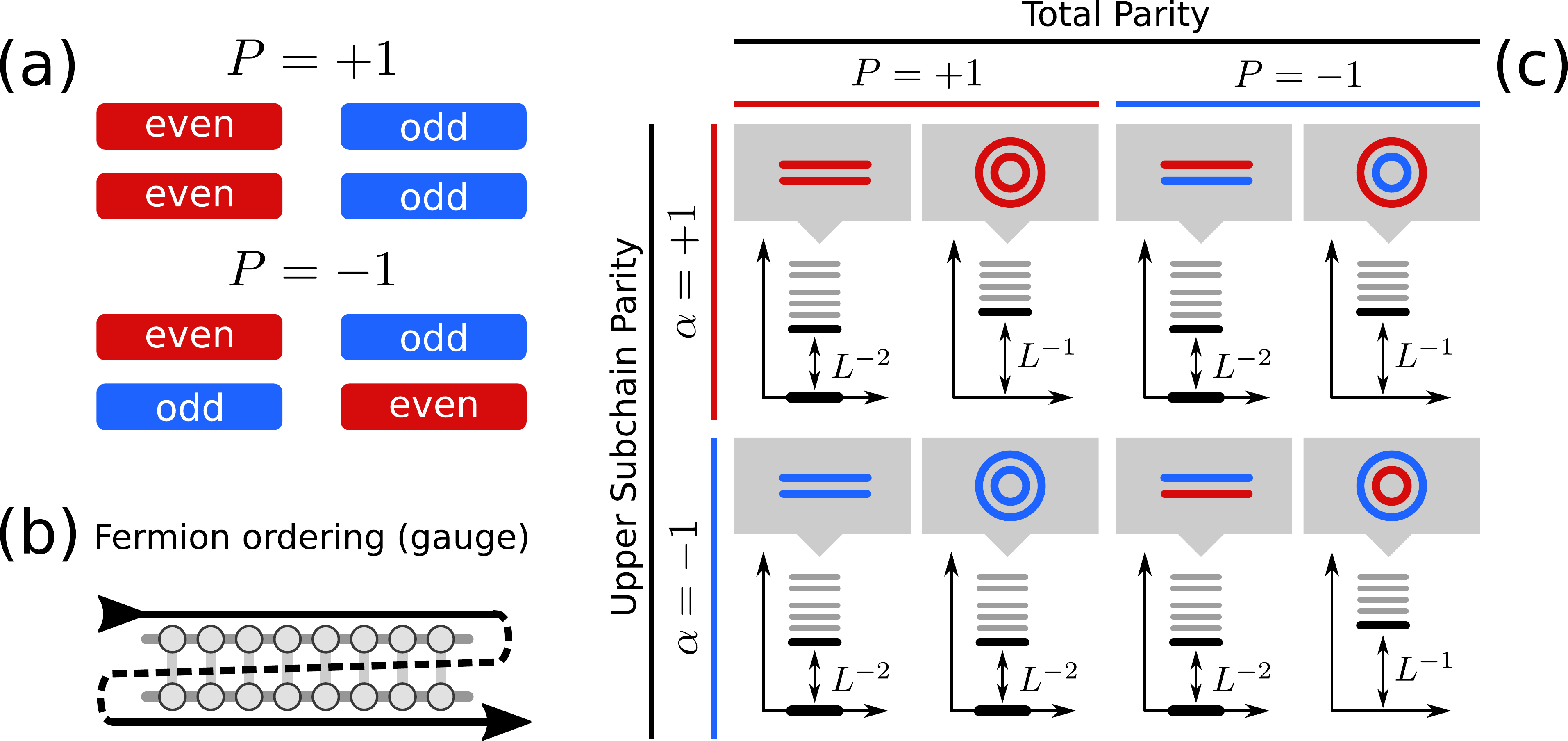}
  \caption{
    \textit{Ground states.}
    (a) For every filling $N$ with parity $P=(-1)^N$, there are two degenerate zero-energy ground states for open boundary conditions, 
    characterized by their (upper) subchain parity $\alpha=P_a$.
    (b) The chosen fermion gauge leads to the simple description of the ground states given in the text.
    (c) Behavior of the spectrum in the low-energy sector of symmetry subspaces classified by the total parity $P$ and the subchain parity $\alpha$.
    Both open (OBC) and periodic (PBC) boundary conditions are shown, zero-energy states are drawn bold.
  }
  \label{fig:5}
\end{figure}

For an \textit{open} ladder, there are exactly two degenerate zero-energy ground states for each filling $0<N<2L$, Fig.~\ref{fig:5}~(a), denoted as
$\Ket{N,\alpha}$ and characterized by the upper chain parity $\alpha\equiv P_a\in\{+1,-1\}$, see the supplemental material for a rigorous proof.
For an appropriate fermion gauge, see Fig.~\ref{fig:5}~(b), each ground state is given by the equal-weight superposition of distributing 
$N$ particles on the two chains constrained by the fixed subchain parity $\alpha$.
To cast this in a formal description, we first introduce the fermion number states $\Ket{\vec n}_x$ with $x=a,b$ and $\vec n\in\{0,1\}^L$, i.e., 
$\Ket{\vec n}_a =(a_{1}^{\dag})^{n_{1}}\!\!\!\!\!\ldots(a_{i}^{\dag})^{n_{i}}\!\!\!\!\!\ldots (a_{L}^{\dag})^{n_{L}}\Ket{0}_a$ 
for the upper chain with the number of fermions $|\vec n|=\sum_{i=1}^L\,n_i$. 
Then the equal-weight superposition states on each chain with a fixed number of particles reduce to
$\Ket{n}_x\equiv \sum_{|\vec n|=n}\,\Ket{\vec n}_x$; note that this state is not normalized.
Finally, the equal-weight superposition with fixed particle number $N$ and subchain-parity $\alpha$ can be written as
\begin{equation}
  \label{eq:5}
  \textstyle\Ket{N,\alpha}=\mathcal{N}_{L,N,\alpha}^{-1/2}\,\sum_{n, (-1)^n=\alpha}\,\Ket{n}_a\Ket{N-n}_b
\end{equation}
where $\mathcal{N}_{L,N,\alpha}^{-1/2}$ is the normalization factor that counts the number of superimposed fermion configurations.

In contrast, for a \textit{closed} ladder the situation is more subtle: For \textit{even} total particle number $N=2K$,
there is a unique zero-energy ground state $\Ket{2K,-1}$ in the odd-odd ($\alpha=-1$) subchain parity sector, 
whereas in the odd-$N$ sectors all states are lifted to finite energy. 
This is summarized in Fig.~\ref{fig:5}~(c) where the low-energy scaling is given as well (see below).

At this point it seems advisable to compare these ground states with those of a {\it single} Majorana chain (Kitaev's chain), 
which in analogy features two zero-energy ground states for open boundary conditions \cite{Kitaev2007}:
For vanishing chemical potential (perfectly localized edge modes), the ground states of the Majorana chain are given by the 
equal-weight superposition of particle number states with fixed (global) parity.  
In contrast, here the chains act as mutual particle reservoirs and the ground state degeneracy arises 
due to two admissible subchain-parity configurations within each fixed particle number sector.

We start exploiting the concise description of the ground states, and derive simple expressions 
for density correlations, superfluid order parameter and the Green's function (single particle correlation).
To this end, it proves useful to define the \textit{parity-split binomial coefficients} (PsBC)
which count the configurations to distribute $N$ particles among $\sum_{i=1}^gL_i$ sites with the 
additional constraint that the parity of subsystem $L_i$ ($1\leq i < g$) is fixed by $\alpha_i=\pm 1$,
\begin{equation}
  \label{eq:6}
  \textstyle\binom{L_1,\dots,L_g}{\alpha_1,\dots,\alpha_{g-1}}_{N}\equiv
  \sum\limits_{n_1,\dots,n_{g-1}}^N\binom{L_g}{N-\sum_{i=1}^{g-1} n_i}\prod_{i=1}^{g-1}\binom{L_i}{n_i} \delta_{n_i}^{\alpha_i}
\end{equation}
with $\delta_{n_i}^{\alpha_i}\equiv\left[1+\alpha_i(-1)^{n_i}\right]/2$. Although we are not aware of simple analytical expressions
(except for special cases, see supplement), the PsBCs can easily be evaluated numerically. Due to the simple structure of the ground states,
all correlation functions and expectation values of $\Ket{N,\alpha}$ can be rewritten in terms of finite combinations of PsBCs.
E.g., the normalization of the two ground states reads $\mathcal{N}_{L,N,\alpha}=\binom{L,L}{\alpha}_N$.

We find that the density-density correlation function factorizes, $\langle x_i^\dag x_{i} y_j^\dag y_{j}\rangle\to \rho^2$ for 
$i\neq j\,;\;x,y\in\{a,b\}$ in the thermodynamic limit $L,N\to\infty$ with fixed particle density $\rho$.
The  pair correlations  read $|\langle x_i^\dag x_{i+1}^\dag y_j y_{j+1}\rangle|\to \rho^2(1-\rho)^2$ for $i\neq j\,;\;x,y\in\{a,b\}$,
and indicate a condensate of $p$-wave pairs with true long-range order.
Note that the results for both correlators do \textit{not} depend on the subchain parity $\alpha$ of the ground states. This is true up to exponential
corrections vanishing with $L\to\infty$. For particularly symmetric setups (e.g., $x\neq y$ and $N$ odd) these corrections even vanish identically.

\begin{figure}[t]
  \centering
  \includegraphics[width=0.95\linewidth]{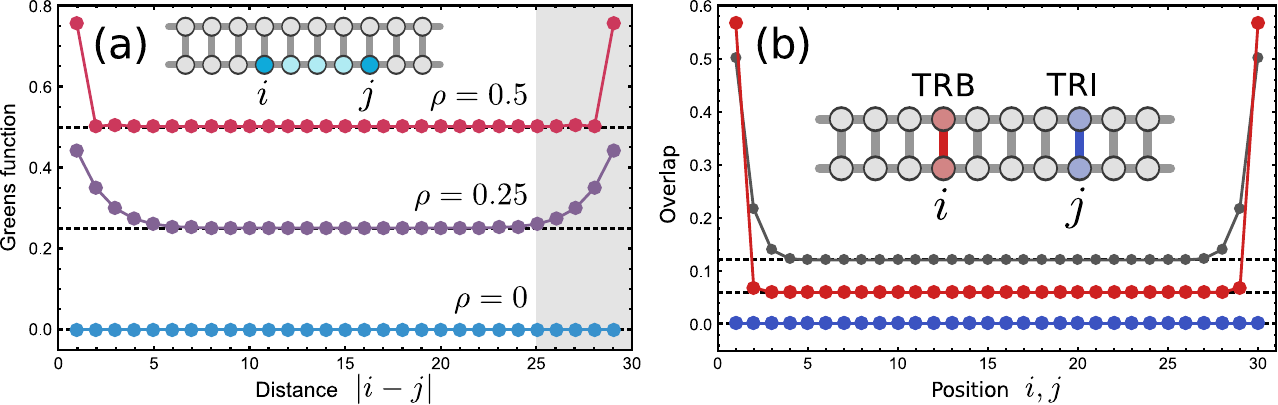}
  \caption{
    \textit{Ground state properties.}
    (a) Intra-chain single particle correlation $\langle a_i^\dag a_j\rangle$ (Green's function) as a function of the distance $|i-j|$ 
    for various fillings $\rho$ and a chain of length $L=30$. The revival for $|i-j|\sim L$ indicates exponentially localized edge states (grey region).
    (b) Overlap of the ground states for time-reversal invariant (TRI) and breaking (TRB) perturbations of $H$ in dependence of the position $i,j$
    of the subchain parity violating single-particle hopping (blue: $\rho=0.5$ TRI, red: $\rho=0.5$ TRB, grey: $\rho=0.25$ TRB). 
  }  
  \label{fig:2}
\end{figure}

The intra-chain Green's function (indicating single particle off-diagonal long-range order \cite{Yang1962}) 
can be expressed in terms of PsBCs ($j>i+1$)
\begin{equation}
  \label{eq:7}
  \textstyle{\langle a_i^\dag a_j\rangle=\mathcal{N}_{L,N,\alpha}^{-1}\,\left[\Lambda_{+1,-\alpha}-\Lambda_{-1,\alpha}\right]}
\end{equation}
where $\Lambda_{\alpha_1,\alpha_2}\equiv\binom{j-i-1,L-j+i-1,L}{\alpha_1,\alpha_2}_{N-1}$.
See the supplement for a detailed derivation.
In the thermodynamic limit one finds exponentially decaying correlations in the bulk, see Fig.~\ref{fig:2}~(a),
\begin{equation}
  \label{eq:8}
  \langle x_i^\dag x_{j} \rangle= e^{-\gamma(\rho){|i-j|}}
  \quad\text{for}\quad 1\ll i,j\ll L;\,x\in\{a,b\}
\end{equation}
where $\gamma$ is some function of the filling with $0 < \gamma(\rho) \leq \infty$ and $\gamma(1/2)=\infty$. 
The boundary terms read $|\langle a_1^\dag a_{L} \rangle| \to \rho(1-\rho)$ in the thermodynamic limit, 
indicating the existence of exponentially localized edge states, Fig.~\ref{fig:2}~(a).

The {\it topological protection} of the ground state degeneracy is most conveniently characterized in terms of their 
indistinguishability by any local perturbation \cite{Nussinov2009,Bonderson2013}. 
Let $\mathcal{O}$ be an arbitrary local (hermitian) operator. Then the expectation values $\Bra{\alpha}\mathcal{O}\Ket{\alpha}$ 
and $\Bra{-\alpha}\mathcal{O}\Ket{-\alpha}$ are identical up to an exponentially small correction ---
as follows from the above analysis of the correlation functions. 
However, for operators violating the subchain-parity $P_x$, also the overlap $\Bra{-\alpha}\mathcal{O}\Ket{\alpha}$
must be taken into account. Then the situation is more subtle. We illustrate this for the simplest case of a single-particle inter-chain 
hopping (the statements can be generalized to more complex $P_x$-violating terms, though). 
Let $\mathcal{O}_j=e^{i\phi}a_j^\dag b_j+e^{-i\phi} b_j^\dag a_j$ with complex hopping phase $\phi\in [0,2\pi)$.
Splitting this perturbation into time-reversal invariant (TRI) and breaking (TRB) contributions, one finds by evaluating the corresponding PsBCs
\begin{subequations}
  \begin{eqnarray}
    \label{eq:9}
    \text{TRI}&:&\;\Bra{-\alpha} a_\delta^\dag b_{\delta}+b_\delta^\dag a_{\delta} \Ket{\alpha}\to 0\\
    \text{TRB}&:&\;\Bra{-\alpha} ia_\delta^\dag b_{\delta}-ib_\delta^\dag a_{\delta} \Ket{\alpha}\to e^{-\mu(\rho){\delta}}
  \end{eqnarray}
\end{subequations}
for the distance $\delta$ from the edges of the ladder, $\delta\ll L$ when $L\to\infty$ and $\rho$ is fixed. 
These site-dependent overlaps are illustrated in Fig.~\ref{fig:2}~(b). Thus the topological ground state degeneracy
for the double wire setup can either be protected by time-reversal symmetry $\mathcal{T}$ or subchain parity $P_x$, and is 
only spoiled if both symmetries are broken at the same time. The latter, however, is not surprising as the two edge states on the upper and lower
wire are not spatially separated. We will show below that our model can be generalized to wire networks, where the different 
edge states become spatially separated. Then it follows immediately that the topological properties
are protected against \textit{any} local operator $\mathcal{O}$ conserving the total number of particles.


\begin{figure}[b]
  \centering
  \includegraphics[width=0.95\linewidth]{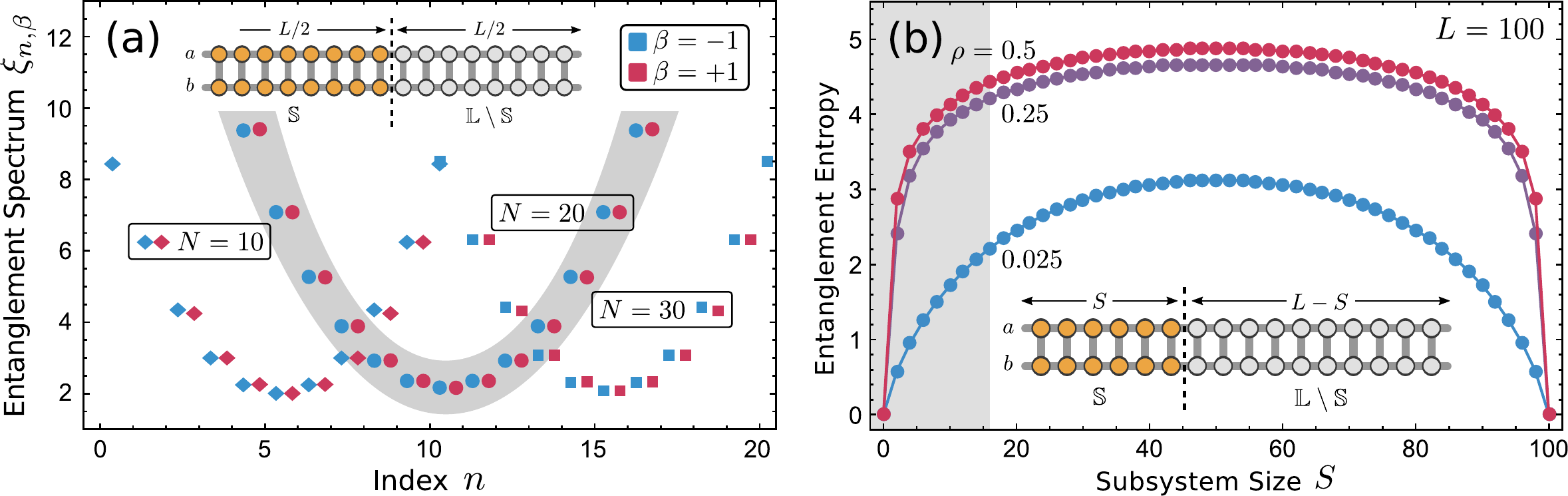}
  \caption{
    \textit{Entanglement.}
    (a) Two branches ($\beta=\pm 1$: red/blue) of the entanglement spectrum for a chain of length $L=20$ and splitting $S=10$
    with fillings $N=10,20,30$ (diamonds, circles, squares). The half-filling branch is highlighted grey. 
    Physically, the index $n$ describes the subsystem filling while $\beta$ describes the subsystem subchain parity. 
    This illustrates the two-fold degeneracy of the entanglement spectrum.
    (b) The entanglement entropy $S^{\text{ent}}$ as a function of subsystem size $S$ for various fillings $\rho$.
    It obeys an area law with logarithmic corrections.
  }
  \label{fig:3}
\end{figure}

{\it Ground state entanglement} ---
Another well-known signature of topological states is a stable degeneracy of the entanglement spectrum (ES) \cite{Li2008,Pollmann2010,Turner2011}.
In our case, the ES of the ground states $\Ket{N,\alpha}$ with respect to a bipartition $\left(\mathbb{S}|\mathbb{L}\setminus\mathbb{S}\right)$ 
of the ladder [see inset of Fig.~\ref{fig:3}~(b)] is given by the Schmidt decomposition
\begin{equation}
  \label{eq:10}
  \Ket{N,\alpha}=\sum_{n}\sum_{\beta=\pm 1}\,e^{-\xi_{n,\beta}/2}\, \Ket{n,\beta}_{\mathbb{S}}\Ket{N-n,\alpha\beta}_{\mathbb{L}\setminus \mathbb{S}}
\end{equation}
and can be written in terms of PsBCs
\begin{equation}
  \label{eq:11}
  \textstyle\xi_{n,\beta}=-\ln{\left[\binom{L-S,L-S}{\alpha\beta}_{N-n}\binom{S,S}{\beta}_{n}/\binom{L,L}{\alpha}_{N}\right]}
\end{equation}
where $\max\{0,N-2L+2S\} \leq n \leq \min\{N,2S\}$ and $\beta=\pm 1$.
The $\beta=\pm 1$-branches of the spectra for a half-split system of length $L=20$ are shown in Fig.~\ref{fig:3}~(a) for different fillings $N$
and reveal the two-fold degeneracy of the ES due to the subsystem subchain parity $\beta$.

In addition, the scaling of the entanglement of a subsystem $\mathbb{S}$ with the environmental system as a function of the subsystem size $S$
in terms of the entanglement entropy $S^\text{ent}[\mathbb S]\equiv-\tr{\rho_{\mathbb S}\ln \rho_{\mathbb S}}$,
with reduced density matrix $\rho_\mathbb{S}=\Tr{\rho}{\mathbb{L}\setminus\mathbb{S}}$, yields insight into the low-energy physics of the theory.
$S^\text{ent}$ can easily be computed from the entanglement spectrum via $S^\text{ent}[\mathbb S]=\sum_{n,\beta}\,e^{-\xi_{n,\beta}}\xi_{n,\beta}$.
Fig.~\ref{fig:3}~(b) shows the (filling dependent) variation of entanglement between a growing subsystem and its environmental system
as function of the subsystem size $S$: It obeys an area law with logarithmic corrections, as expected from a critical (gapless) 1D system.
That is, in contrast to the gapped Majorana chain, here we face a low-energy theory of gapless Goldstone modes due to particle number conservation.
With this in mind, we have a closer look at the low-energy excitations.


{\it Low-energy excitations} ---
The single-chain Hamiltonians $H^x$ for an open ladder can be mapped to the ferromagnetic, isotropic Heisenberg chain
via a Jordan-Wigner transformation. The complete spectrum of $H^x$ is therefore accessible via the Bethe-Ansatz \cite{Bethe1931}.
Exploiting this mapping, it is possible to construct the analog of single magnon states for our theory.  
These exact low energy eigenstates for the open \textit{double}-chain take the form
\begin{equation}
  \label{eq:13}
  \Ket{k; N,\alpha}= P_1^a(k)\oplus P_1^b(k)\,\Ket{N,\alpha}
\end{equation}
with momentum $k=m\frac{\pi}{L}$, $0\leq m < L$, and the operator
\begin{equation}
  \label{eq:12}
  \textstyle P_1^x(k)=\sum_{j=1}^L\,\cos\left[\frac{k}{2}(2j-1)\right](-1)^{x_j^\dag x_j}\,.
\end{equation}
The eigenenergies are given by a quadratic excitation spectrum $E_k=4\sin^2\frac{k}{2}$.  
This  behavior of the Goldstone mode is in excellent agreement with the appearance of a 
true condensate and vanishing compressibility; recall that for any fixed number of particles 
there is a zero-energy ground state. An equivalent behavior is well-known for non-interacting bosons
and the ferromagnetic Heisenberg model in one-dimension. The interpretation of these features is that 
our model is exactly solvable at a critical point.


{\it Wire networks and non-abelian statistics} ---
A crucial aspect of our model is that the derivation of the exact zero-energy ground states can be 
straightforwardly generalized to much more complicated wire networks consisting of open and 
closed single chains sectionally connected to ladder segments with arbitrary positive coupling strengths, 
see Fig.~\ref{fig:4}~(a) for an example; the general formalism is presented in the supplement. 
It follows immediately that the ground state degeneracy scales as $2^{E/2-1}$ with $E\geq 2$ the number of open subchain ends. 
This scaling is in agreement with the interpretation of the localized edge states as interacting equivalent of Majorana zero modes. 
In order to provide a rigorous proof of the topological properties characterizing the localized edge states, we derive the full braiding statistics. 
Note that the gap $\Delta$ closes algebraically, $\Delta \propto 1/L^2$, Fig.~\ref{fig:5}~(c). 
This still allows for a generalized notion of braiding and thereby probing the edge state statistics \cite{Bonderson2013}.

\begin{figure}[t]
  \centering
  \includegraphics[width=1.0\linewidth]{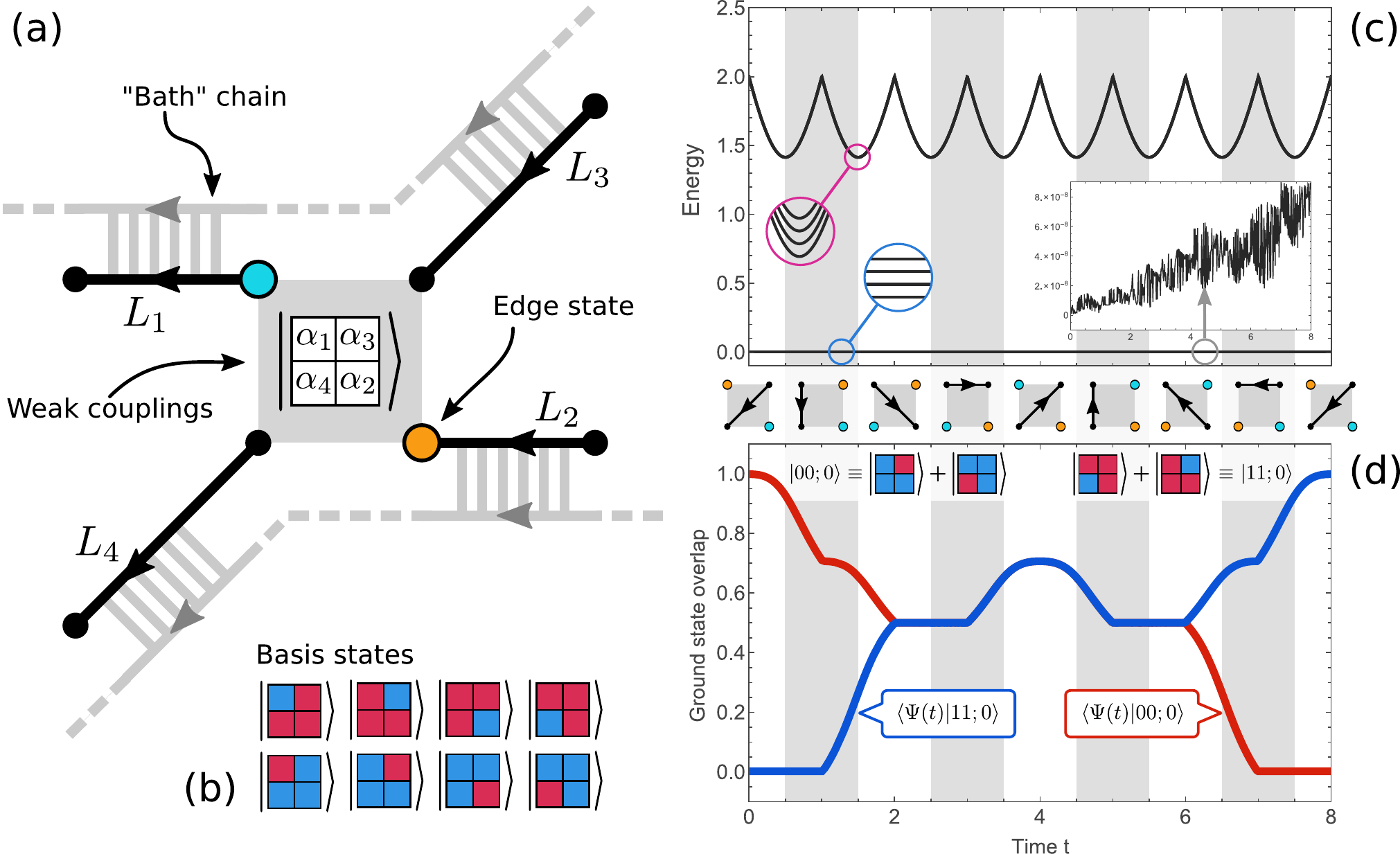}
  \caption{
    \textit{Braiding.}
    (a) Setup of four open chains $L_i$, $i=1,2,3,4$ (black) with controllable weak single-particle couplings between the inner four endpoints.
    The partner chains (grey) are not involved in the braiding and can be disregarded.
    (b) The dynamics takes place in the $8$-dimensional Hilbert space spanned by the subchain-parity eigenstates with fixed total parity
    $\alpha=\alpha_1\alpha_2\alpha_3\alpha_3=-1$. The colors denote the subchain parities $\alpha_i$ of the four black chains.
    (c) Spectrum of the weak coupling Hamiltonian during the braiding procedure depicted below the plot. A black arrow indicates
    single particle hopping. There are four degenerate zero-energy ground states. The deviation from zero-energy (perfect adiabaticity) 
    due to the finite time evolution is shown in the inset ($\sim 10^{-8}$).
    (d) Time evolution for the initial zero-energy state $\Ket{00;0}$. Shown are the (moduli of the) overlaps with $\Ket{00;0}$ and $\Ket{11;0}$
    (see inset). 
  }
  \label{fig:4}
\end{figure}

In order to braid two localized edge states, we consider the wire network of four open subchains coupled by a common ``bath'' chain
depicted in Fig.~\ref{fig:4}~(a) and described by $H_0$. Only the highlighted chains $L_i$ ($i=1,\dots,4$) take part in the braiding evolution. 
Thus the greyed out subchain can henceforth be neglected and considered as a ``bath'', the effect of which is fully incorporated 
into the exactly known ground states.
Note that the zero-energy states of the uncoupled subchains are given by the total filing $N$ and the subchain parities 
$\alpha_1,\dots,\alpha_4$, spanning a $2^4=16$ dimensional ground state space in each particle number sector.
As we are only considering interactions between the four subchains, the total subchain parity $\alpha=\prod_i\alpha_i$ is conserved and may be
fixed at $\alpha=-1$, reducing the number of relevant ground states to eight, see Fig.~\ref{fig:4}~(b). 
The braiding of the edges states is described by $H_\text{int}(t)$ and achieved by adiabatically turning off the coupling between two edges and turning on the coupling between 
the next two edges; the full sequence of couplings for the winding of two edge states around each other is shown below Fig.~\ref{fig:4}~(c), 
where arrows indicate single-particle couplings analogous to $A_i(\mathds{1}+A_i)$. 

The analysis is performed by the full numerical time evolution of the Hamiltonian $ H(t)=H_0+\varepsilon/L^2\,H_\text{int}(t)$
with $\varepsilon \ll 1$ and $0\leq t \leq 8$ to guarantee the (quasi)-adiabatic evolution.  
Starting with the initial zero-energy state $\Ket{00;0}$, Fig.~\ref{fig:4}~(d), characterized 
by $\alpha_1=-1=\alpha_2$ and $\alpha_3\alpha_4=-1$, yields the orthogonal final 
state $\Ket{11;0}=\exp\left[-i\int\mathrm{d}t\,H_\text{int}(t)\right]\Ket{00;0}$, characterized 
by $\alpha_1=+1=\alpha_2$ and $\alpha_3\alpha_4=-1$. 
Repeating the analysis for alternative braiding operations, we find the non-abelian holonomy acting on the degenerate 
ground state space that qualifies the edge states as Ising anyons \cite{1_kitaev_AESMB_S}, 
which corresponds to the braiding statistics of Majorana edge modes in non-interacting theories.


{\it Conclusion} ---
We presented a microscopic model of interacting fermions giving rise to a gapless topological state with
non-abelian edge states. The system is at a critical point and certain perturbations to the Hamiltonian will drive 
the system into a phase separated state (e.g., increasing the attractive interactions), while we expect resilience 
of the topological properties against other perturbations (e.g., increasing the hopping). Then the ground state should
be well described by an approach based on bosonization similar to \cite{Cheng2011,Fidkowski2011a,Sau2011,Ruhman,Keselman2015}, 
and might be connected to the state studied with DMRG~\cite{Kraus2013a}.


{\it Note added} ---
During the final steps of preparation, we became aware of related results studied by 
Iemini~\textit{et al.}~\footnote{F. Iemini, L. Mazza, D. Rossini, S. Diehl, and R. Fazio, \textit{in preparation}.}.


\begin{acknowledgments}
{\it Acknowledgements} ---
We acknowledge support by the Deutsche Forschungsgemeinschaft (DFG) within SFB/TRR~21.
H.P.B thanks Ehud Altman for his hospitality at the Weizmann Institute.
\end{acknowledgments}


\bibliographystyle{apsrev4-1}
\bibliography{bibliography}

\clearpage


\section*{Supplementary material}

Nicolai Lang[*] and Hans Peter Büchler
*Institute for Theoretical Physics III, University of Stuttgart, 70550 Stuttgart, Germany*
(Dated: April 16, 2015)

**Derivation of zero-energy ground states**

It is straightforward to show that the two states $|N, \alpha\rangle$, $\alpha = \pm 1$, in each filling sector $N$ are annihilated by $H$, and therefore belong to the ground state space. Here we sketch how to derive their uniqueness by constructing the zero-energy eigenstates of $H$ from scratch. To this end, consider the binary vectors $(\boldsymbol{n}, \boldsymbol{m}) \in \{0, 1\}^{2L}$ describing the fermion configurations on the two-leg ladder. The Fock space is spanned by number states $|\boldsymbol{n}, \boldsymbol{m}\rangle \equiv |\boldsymbol{n}\rangle_a |\boldsymbol{m}\rangle_b \in \mathcal{H}$, where we choose the fermion ordering (gauge) according to Fig. 1 (a); this defines a representation of the fermion algebra $\{a_i, a_i^\dagger, b_i, b_i^\dagger\}_{1 \leq i \leq L}$ on $\mathcal{H}$.

The action of this representation allows us to write

$$A_i^x(\mathbb{1} + A_i^x)|\boldsymbol{n}, \boldsymbol{m}\rangle = |\boldsymbol{n}, \boldsymbol{m}\rangle - |\varepsilon_i^x(\boldsymbol{n}, \boldsymbol{m})\rangle \quad (1)$$

and

$$B_i(\mathbb{1} + B_i)|\boldsymbol{n}, \boldsymbol{m}\rangle = |\boldsymbol{n}, \boldsymbol{m}\rangle - |\tau_i(\boldsymbol{n}, \boldsymbol{m})\rangle \quad (2)$$

where we used the single-bit swap-operators

$$\varepsilon_i^a((\ldots, n_i, n_{i+1}, \ldots), \boldsymbol{m}) = ((\ldots, n_{i+1}, n_i, \ldots), \boldsymbol{m}) \quad (3a)$$
$$\varepsilon_i^b(\boldsymbol{n}, (\ldots, m_i, m_{i+1}, \ldots)) = (\boldsymbol{n}, (\ldots, m_{i+1}, m_i, \ldots)) \quad (3b)$$

for arbitrary $(\boldsymbol{n}, \boldsymbol{m})$ and the bit-pair swap-operators

$$\tau_i((\ldots, n_i, n_{i+1}, \ldots), (\ldots, m_i, m_{i+1}, \ldots)) = ((\ldots, m_i, m_{i+1}, \ldots), (\ldots, n_i, n_{i+1}, \ldots)) \quad (4)$$

for $(\boldsymbol{n}, \boldsymbol{m})$ with $n_i = n_{i+1}$ and $m_i = m_{i+1}$, and $\tau_i = \text{Id}$ otherwise.

It is convenient to introduce the partition $\bigcup_N B_N = \{0, 1\}^{2L}$ of binary vectors with the decomposition $B_N = B_N^+ \cup B_N^-$, where $B_N^\alpha$ is characterized by vectors $(\boldsymbol{n}, \boldsymbol{m})$ with fixed filling $\sum_i (n_i + m_i) = N$, and fixed sub-parity $\prod_i (-1)^{n_i} = \alpha$. It is now easy to see that (i) $\varepsilon_i^x, \tau_i : B_N^\alpha \to B_N^\alpha$ are bijections for $x = a, b$ and $1 \leq i \leq L$ (note that $\varepsilon_i^x \varepsilon_i^x = \text{Id} = \tau_i \tau_i$), and (ii) for any pair $(\boldsymbol{n}, \boldsymbol{m}), (\boldsymbol{n}', \boldsymbol{m}') \in B_N^\alpha$ there is a finite sequence $\sigma$ of $\varepsilon_i^x, \tau_i$ such that $(\boldsymbol{n}, \boldsymbol{m}) = \sigma(\boldsymbol{n}', \boldsymbol{m}')$. That is, the connected components of $\{0, 1\}^{2L}$ under the described family of bit-operations $\Pi \equiv \{\pi\} = \{\varepsilon_i^a, \varepsilon_i^b, \tau_i\}_{1 \leq i \leq L}$ are exactly the $B_N^\alpha$ for $N = 0, \ldots, 2L$ and $\alpha = \pm 1$.

We proceed with a generic state $|\Psi\rangle \in \mathcal{H}$,

$$|\Psi\rangle = \sum_{(\boldsymbol{n}, \boldsymbol{m})} \Psi(\boldsymbol{n}, \boldsymbol{m}) |\boldsymbol{n}, \boldsymbol{m}\rangle \in \mathcal{H} \quad (5)$$

and evaluate the energy expectation value

$$\langle \Psi | H | \Psi \rangle = \sum_{i,x} \langle \Psi | A_i^x (\mathbb{1} + A_i^x) | \Psi \rangle + \sum_i \langle \Psi | B_i (\mathbb{1} + B_i) | \Psi \rangle \quad (6)$$

which reads

$$\langle \Psi | H | \Psi \rangle = \sum_{\pi \in \Pi} \sum_{(\boldsymbol{n}, \boldsymbol{m}), (\boldsymbol{n}', \boldsymbol{m}')} \Psi^*(\boldsymbol{n}', \boldsymbol{m}') \Psi(\boldsymbol{n}, \boldsymbol{m}) \langle \boldsymbol{n}', \boldsymbol{m}' | [|\boldsymbol{n}, \boldsymbol{m}\rangle - |\pi(\boldsymbol{n}, \boldsymbol{m})\rangle] \quad (7a)$$
$$= \sum_{\pi \in \Pi} \sum_{(\boldsymbol{n}, \boldsymbol{m})} [\Psi^*(\boldsymbol{n}, \boldsymbol{m}) \Psi(\boldsymbol{n}, \boldsymbol{m}) - \Psi^* \pi(\boldsymbol{n}, \boldsymbol{m}) \Psi(\boldsymbol{n}, \boldsymbol{m})]. \quad (7b)$$

Doubling the sum, substituting $\pi(\boldsymbol{n}, \boldsymbol{m}) \to (\boldsymbol{n}, \boldsymbol{m})$ and using the bijectivity, $\pi\{0, 1\}^{2L} = \{0, 1\}^{2L}$, yields

$$\langle \Psi | H | \Psi \rangle = \frac{1}{2} \sum_{\pi \in \Pi} \sum_{(\boldsymbol{n}, \boldsymbol{m})} [\Psi^*(\boldsymbol{n}, \boldsymbol{m}) \Psi(\boldsymbol{n}, \boldsymbol{m}) - \Psi^* \pi(\boldsymbol{n}, \boldsymbol{m}) \Psi(\boldsymbol{n}, \boldsymbol{m})] \quad (8a)$$
$$+ \frac{1}{2} \sum_{\pi \in \Pi} \sum_{(\boldsymbol{n}, \boldsymbol{m})} [\Psi^* \pi(\boldsymbol{n}, \boldsymbol{m}) \Psi \pi(\boldsymbol{n}, \boldsymbol{m}) - \Psi^*(\boldsymbol{n}, \boldsymbol{m}) \Psi \pi(\boldsymbol{n}, \boldsymbol{m})] \quad (8b)$$
$$= \frac{1}{2} \sum_{\pi \in \Pi} \sum_{(\boldsymbol{n}, \boldsymbol{m})} |\Psi(\boldsymbol{n}, \boldsymbol{m}) - \Psi \pi(\boldsymbol{n}, \boldsymbol{m})|^2. \quad (8c)$$

Splitting the second sum into connected components gives the final result,

$$\langle\Psi|H|\Psi\rangle = \tfrac{1}{2} \sum_{N,\alpha} \sum_{\pi\in\Pi} \sum_{(\boldsymbol{n},\boldsymbol{m})\in B_N^\alpha} |\Psi(\boldsymbol{n},\boldsymbol{m}) - \Psi\pi(\boldsymbol{n},\boldsymbol{m})|^2 . \qquad (9)$$

Since $|\Psi\rangle$ is a zero-energy ground state if and only if $\langle\Psi|H|\Psi\rangle = 0$, and the $B_N^\alpha$ are the connected components under the action of $\Pi$, we conclude that for any normalizable zero-energy state it is necessary and sufficient that the amplitudes are constant within each sector $B_N^\alpha$,

$$H|\Psi\rangle = 0 \Leftrightarrow \Psi(\boldsymbol{n},\boldsymbol{m}) = \Psi_N^\alpha, \ \forall (\boldsymbol{n},\boldsymbol{m}) \in B_N^\alpha. \qquad (10)$$

These are exactly the equal-weight superpositions $|N,\alpha\rangle$ of fixed particle number $N$ and subchain parity $\alpha$ introduced in the main text. We stress that these statements are modified for closed boundary conditions within the even-even and odd-even/even-odd sectors where some of the summands in Eq. (9) read $|\Psi(\boldsymbol{n},\boldsymbol{m}) + \Psi\pi(\boldsymbol{n},\boldsymbol{m})|^2$ due to the fermionic statistics. From this it follows immediately that no normalizable state with exact zero-energy can exist, though the ground state energy may vanish in the thermodynamic limit nevertheless (see below).

### Parity-split binomial coefficients

As explained in the main text, the evaluation of arbitrary expectation values and correlators of the zero-energy ground states $|N,\alpha\rangle$ is particularly efficient due to their simple structure as equal-weight superposition of number states with fixed total particle number $N$ and (upper) subchain parity $\alpha$. The evaluation is then reduced to counting fermion configurations with specific parity constraints $\alpha_1,\ldots,\alpha_{g-1}$ on $g-1$ subsystems $L_1,\ldots,L_{g-1}$ (the $g$th subsystem is the "rest" or "environment"). Therefore we introduced the parity-split binomial coefficients (PsBC)

$$\binom{L_1,\ldots,L_g}{\alpha_1,\ldots,\alpha_{g-1}}_N \equiv \sum_{n_1,\ldots,n_{g-1}}^N \binom{L_g}{N-\sum_{i=1}^{g-1} n_i} \prod_{i=1}^{g-1} \delta_{n_i}^{\alpha_i}\binom{L_i}{n_i} \qquad (11)$$

with $\delta_{n_i}^{\alpha_i} \equiv [1 + \alpha_i(-1)^{n_i}]/2$. They can be evaluated efficiently on a computer, though we are not aware of closed forms for arbitrary splittings $g$. However, the (most important) PsBC used for the normalizing factor and describing the splitting of the double chain in upper and lower wires of size $L$, $\mathcal{N}_{L,N,\alpha} = \binom{L,L}{\alpha}_N$, allows for a closed form, namely

$$\binom{L,L}{\alpha}_N = \tfrac{1}{2}\binom{2L}{N} + \tfrac{\alpha}{2} \tfrac{1+(-1)^N}{2}(-1)^{N/2}\binom{L}{N/2}. \qquad (12)$$

Thus the leading order is $\tfrac{1}{2}\binom{2L}{N}$. This is exact for odd $N$, as one would expect due to the chain-exchange symmetry relating the even-odd and odd-even sectors. For even fillings $N$, however, there are corrections of order $\binom{L}{N/2}$ due to the inequivalence of the even-even and odd-odd sectors.

As a final remark, we give the following alternative form of a generic $g=2$ split PsBC

$$\binom{L_1,L_2}{\alpha}_N = \tfrac{1}{2}\binom{L_1+L_2}{N} + \tfrac{\alpha}{2} \sum_{n=0}^N (-1)^n \binom{L_1}{n}\binom{L_2}{N-n} \qquad (13)$$

which separates the leading contribution from the $\alpha$-dependent corrections. We make use of this result below.

### Evaluation of correlators

To exemplify the evaluation of correlators, we show how the density-density correlations, the suprafluid pair correlations, and the intra-chain Green's function can be evaluated in terms of PsBCs.

We start with the density-density correlations for which we find

$$\langle x_i^\dagger x_i y_j^\dagger y_j\rangle = \begin{cases} \binom{L-1,L-1}{-\alpha}_{N-2}\binom{L,L}{\alpha}_N^{-1} & \text{for } x \neq y \\ \binom{L-2,L}{\alpha}_{N-2}\binom{L,L}{\alpha}_N^{-1} & \text{for } x = y \end{cases}.$$

This follows from counting all fermion configurations in $|N,\alpha\rangle$ with occupied sites $i$ and $j$. Along the same lines, counting configurations with occupied sites $j$ and $j+1$ and vacant sites $i$ and $i+1$ yields similar expressions for the suprafluid pair correlations,

$$|\langle x_i^\dagger x_{i+1}^\dagger y_j y_{j+1}\rangle| = \begin{cases} \binom{L-2,L-2}{\alpha}_{N-2}\binom{L,L}{\alpha}_N^{-1} & \text{for } x \neq y \\ \binom{L-4,L}{\alpha}_{N-2}\binom{L,L}{\alpha}_N^{-1} & \text{for } x = y \end{cases}.$$

In the light of these two correlators, it seems advisable to consider the general expression

$$\binom{L-l_1,L-l_2}{\alpha}_{N-m}\binom{L,L}{\beta}_N^{-1}$$

in more detail. Using Eq. (13) from above and setting $N = 2L\rho$, one can derive the expression

$$\binom{L-l_1,L-l_2}{\alpha}_{2L\rho-m}\binom{L,L}{\beta}_{2L\rho}^{-1} = \binom{2L-l_1-l_2}{2L\rho-m}\binom{2L}{2L\rho}^{-1} \qquad (14)$$
$$+ \alpha\,\mathcal{O}(e^{-\lambda L}) + \beta\,\mathcal{O}(e^{-\mu L})$$

which shows that any distinction due to the subparities $\alpha,\beta$ vanishes exponentially. Straightforward simplifications for $L\to\infty$ lead to the final result

$$\binom{L-l_1,L-l_2}{\alpha}_{2L\rho-m}\binom{L,L}{\beta}_{2L\rho}^{-1} \qquad (15)$$
$$\sim \rho^m(1-\rho)^{l_1+l_2-m} + \alpha\,\mathcal{O}(e^{-\lambda L}) + \beta\,\mathcal{O}(e^{-\mu L})$$


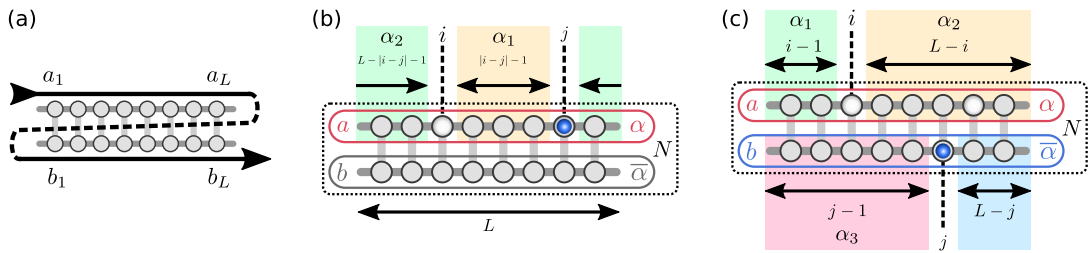

Figure 1. *Green's function.* (a) Fermion ordering (gauge) defining the number state basis used to construct the ground states. (b) Derivation of the single particle intra-chain Green's function of the zero-energy ground states $|N, \alpha\rangle$ by counting states of fixed subsector parities $\alpha_1$ and $\alpha_2$. The single particle hopping from site $j$ to the (vacant) site $i$ measures the parity $\alpha_1$ of the subsector in between (yellow). (c) The partition used for the derivation of the inter-chain single-particle overlap $\langle N, -\alpha | a_i^\dagger b_j | N, \alpha\rangle$. The single particle hopping between site $j$ and the (vacant) site $i$ measures the double-segment (yellow and red) parity $\alpha_2 \alpha_3$ due to the fermion gauge shown in (a). The results in both cases can be expressed in terms of PsBCs and are described in the text.

The correlators from above follow now as special cases, $\langle x_i^\dagger x_i y_j^\dagger y_j \rangle = \rho^2$ and $|\langle x_i^\dagger x_{i+1}^\dagger y_j y_{j+1}\rangle| = \rho^2 (1-\rho)^2$, up to exponential corrections depending on the subchain parity $\alpha$ of the ground state $|N, \alpha\rangle$.

To conclude this section, we focus on the intra-chain Green's function. Here the derivation is more subtle since the parity of the single-chain segment separating the two support regions of the operators must be taken into account. For $j > i$ one finds

$$\langle N, \alpha | a_i^\dagger a_j | N, \alpha \rangle = \binom{L, L}{\alpha}_N^{-1} [\Lambda_{+1, -\alpha} - \Lambda_{-1, \alpha}] \quad (16)$$

where

$$\Lambda_{\beta_1, \beta_2} = \binom{j - i - 1, L - j + i - 1, L}{\beta_1, \beta_2}_{N-1} \quad (17)$$

is a $g = 3$ split PsBC. This follows from simple counting arguments and the convenient partition shown in Fig. 1 (b): $a_i^\dagger a_j$ chooses all states of the equal-weight superposition in $|N, \alpha\rangle$ with vacant site $i$ and occupied site $j$. These states are not annihilated and contribute $\pm 1$ to the result as there is exactly one bra state in $\langle N, \alpha|$ for each number state in $a_i^\dagger a_j |N, \alpha\rangle$. The sign of the contribution depends on the parity of the segment between site $i$ and $j$ due to our fermion ordering from left to right along the chains, Fig. 1 (a). We therefore need to count all configurations of the $N - 1$ unaffected fermions (one is fixed by the hopping) on $g = 3$ segments:

1. Segment: length $|i - j| - 1$ and parity $\alpha_1$ between sites $i$ and $j$.

2. Segment: the rest of the upper chain of length $L - |i - j| - 1$ (excluding sites $i$ and $j$) with parity $\alpha_2$.

3. Segment: the lower chain of length $\alpha$ with fixed parity $\bar{\alpha} = \alpha(-1)^N$.

The total parity of the first two segments is fixed to $\alpha_1 \alpha_2 = -\alpha$ since a single fermion is excluded and used for the hopping. Therefore we find exactly $\Lambda_{+1, -\alpha}$ configurations which contribute $+1$ (the parity of the segment between $i$ and $j$ is $+1$) and $\Lambda_{-1, \alpha}$ configurations which contribute $-1$. Divided by the total number of configurations in $|N, \alpha\rangle$, this yields the result presented above.

A numerical evaluation is given in Fig. 3 (a) of the main text and reveals the single particle off-diagonal long range order close to the boundaries of the ladder due to the fixed subchain parity.

**Inter-chain single-particle hopping**

We extend the previous section by a detailed derivation of the subchain parity violating single-particle hopping between adjacent chains. As already explained in the main text, it is convenient to split the the generic perturbation

$$\mathcal{O}_j = e^{i\phi} a_j^\dagger b_j + e^{-i\phi} b_j^\dagger a_j \quad (18)$$

with $\phi \in [0, 2\pi)$ into time-reversal breaking (TRB) and time-reversal invariant (TRI) contributions,

$$\text{TRI} : h_j^+ = a_j^\dagger b_j + b_j^\dagger a_j \quad (19a)$$
$$\text{TRB} : h_j^- = a_j^\dagger b_j - b_j^\dagger a_j . \quad (19b)$$

We are interested in overlaps $\langle -\alpha | h_j^\pm | \alpha \rangle$ where we omit the invariant total particle number in the naming scheme of the ground states. This reduces to the computation of the inter-chain single-particle correlators $\langle -\alpha | a_j^\dagger b_j | \alpha \rangle$ and $\langle -\alpha | b_j^\dagger a_j | \alpha \rangle$.

Splitting the double chain into four segments, as sketched in Fig. 1 (c), allows an evaluation along the same lines described above for the intra-chain Green's function:


















Counting fermion configurations with one of the sites $i, j$ empty and the other occupied, sorting them according to their subsegment parities $\alpha_i$, $i = 1, 2, 3$, yields the expression

$$\langle -\alpha | a_i^\dagger b_j | \alpha \rangle = \binom{L,L}{\alpha}_N^{-1/2} \binom{L,L}{-\alpha}_N^{-1/2} \left( \Lambda_{\alpha,+1,+1} + \Lambda_{-\alpha,-1,-1} - \Lambda_{\alpha,+1,-1} - \Lambda_{-\alpha,-1,+1} \right) \tag{20}$$

with the $g = 4$ split PsBCs

$$\Lambda_{\alpha_1,\alpha_2,\alpha_3} = \binom{i-1,L-i,j-1,L-j}{\alpha_1,\alpha_2,\alpha_3}_{N-1}. \tag{21}$$

The special case $i = j$ allows us to write

$$\langle -\alpha | h_j^\pm | \alpha \rangle \propto \left( \Lambda_{\alpha,+1,+1} + \Lambda_{-\alpha,-1,-1} - \Lambda_{\alpha,+1,-1} - \Lambda_{-\alpha,-1,+1} \right) \pm \left( \Lambda_{\alpha,-1,-1} + \Lambda_{-\alpha,+1,+1} - \Lambda_{-\alpha,+1,-1} - \Lambda_{\alpha,-1,+1} \right) \tag{22}$$

for the Hamiltonian perturbation overlaps. Doing the math yields for the time-reversal invariant (TRI) perturbation

$$\langle -\alpha | h_j^+ | \alpha \rangle \propto \sum_{n_1,n_2,n_3=0}^{N-1} \binom{L-j}{N-1-n_1-n_2-n_3} \binom{j-1}{n_1} \binom{L-j}{n_2} \binom{j-1}{n_3} (-1)^{n_2+n_3} = \frac{1-(-1)^N}{2} (-1)^{\frac{N-1}{2}} \binom{L-1}{(N-1)/2}. \tag{23}$$

Incorporating the normalization factors yields

$$\langle -\alpha | h_j^+ | \alpha \rangle = \left[ 1 - (-1)^N \right] (-1)^{\frac{N-1}{2}} \binom{L-1}{(N-1)/2} \binom{2L}{N}^{-1} \sim \left[ 1 - (-1)^N \right] (-1)^{\frac{N-1}{2}} \mathcal{O}\left( e^{-\lambda(\rho)L} \right) \tag{24}$$

for $N = 2\rho L$, $L \to \infty$ where we used expression (12) for the normalization factors. Therefore we find that the overlap due to TRI perturbations vanishes identically for even fillings $N$, whereas in odd filling sectors exponentially vanishing corrections occur.

Finally, an analogous calculation for the time-reversal breaking (TRB) hopping yields

$$\langle -\alpha | h_j^- | \alpha \rangle \propto \alpha \sum_{n_1,n_2,n_3=0}^{N-1} \binom{L-j}{N-1-n_1-n_2-n_3} \binom{j-1}{n_1} \binom{L-j}{n_2} \binom{j-1}{n_3} (-1)^{n_1+n_3} = \alpha \sum_{n=0}^{N-1} (-1)^n \binom{2j-2}{n} \binom{2L-2j}{N-1-n} \tag{25}$$

which then gives rise to the lengthy expression

$$\langle -\alpha | h_j^- | \alpha \rangle = 2\alpha \left[ \binom{2L}{N}^2 - \frac{1+(-1)^N}{2} \binom{L}{N/2}^2 \right]^{-1/2} \sum_{n=0}^{N-1} (-1)^n \binom{2j-2}{n} \binom{2L-2j}{N-1-n}. \tag{26}$$

A numerical evaluation was given in Fig. 3 (b) of the main text and shows for fixed filling $\rho$ and $L \to \infty$ exponentially decaying overlaps in the bulk (the localization length depending on $\rho$, vanishing at half-filling) but finite ones for $j \to 1, L$. This is derived easily in the odd filling sector where for $j = 1, L$ one finds

$$\left| \langle -\alpha | h_{1,L}^- | \alpha \rangle \right| = 2 \binom{2L-2}{N-1} \binom{2L}{N}^{-1} \sim 2\rho(1-\rho) \quad \text{for} \quad L \to \infty \tag{27}$$

which becomes $1/2$ for half filling, $\rho = 0.5$.

The vanishing overlaps for TRI and the finite, edge-localized overlaps for TRB perturbations are analogous to the fusion of Majorana fermions in a parallel two-wire setup of non-interacting Majorana chains due to time-reversal symmetry breaking couplings — which illustrates the breakdown of a $\mathbb{Z}$ topological index in symmetry class BDI to the $\mathbb{Z}_2$ index of D [1, 2].

**Low-energy sector**

Here we explain the gap and ground state energy scalings depicted in Fig. 2 (c) of the main text in more detail. We start with the algebraic energy gap closing $\Delta E_0(L) \propto L^{-2}$ in all four parity sectors (total parity $(-1)^N = \pm 1$, upper subchain parity $\alpha = \pm 1$) for *open* boundary conditions. In this case, there is a unique zero-energy ground state $|N, \alpha\rangle$ in each sector and we ask about the energy of the first excited state defining the energy gap $\Delta E_0(L)$. As explained in the main text, one can derive the single particle excitations in each sector $(N, \alpha)$ rigorously. The lowest lying single-particle excitation has momentum $k = \frac{\pi}{L}$ and energy

$$E_1(L) = 4\sin^2 \frac{\pi}{2L} \sim \frac{\pi^2}{L^2} \tag{28}$$

which yields an upper bound for the energy gap,

$$\Delta E_0(L) \lesssim \frac{\pi^2}{L^2}. \tag{29}$$

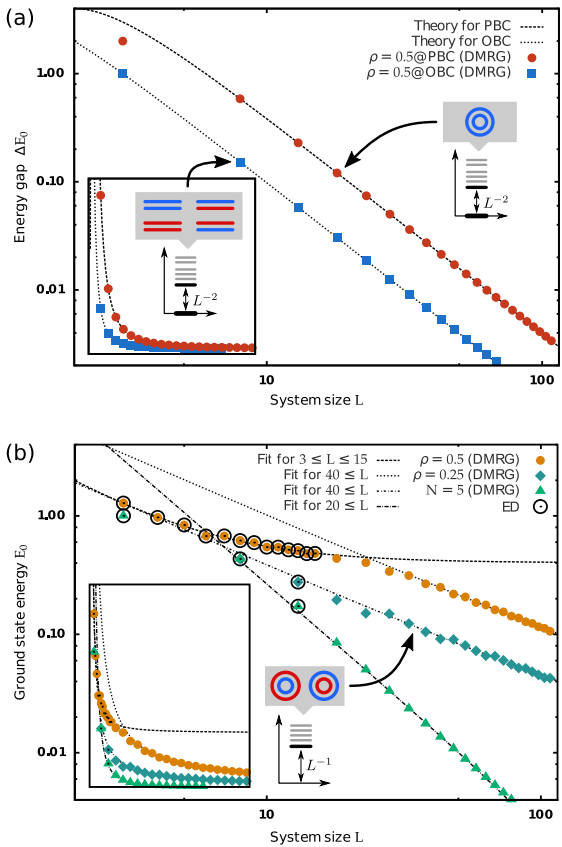

Figure 2. *DMRG Results.* (a) Log-log plot of the energy gap $\Delta E_0$ scaling for $L \to \infty$ for open and closed ladders in the even particle number sector within the odd-odd subsector for $\rho = 0.5$. Comparison with the analytic result for single-magnon excitations with momentum $k = \pi/L$ (OBC) and $k = 2\pi/L$ (PBC) shows perfect agreement. The insets show the same data in linear scale. Note that for OBC the results can be transferred to the other subchain parity sectors. (b) Log-log plot of the ground state energy $E_0$ scaling for $L \to \infty$ for a closed ladder in the odd particle number sector for both fixed filling $\rho = 0.5, 0.25$ and particle number $N = 5$. For fixed filling, the ground state energy vanishes with $1/L$, for fixed particle number with $1/L^2$. Note that exact diagonalization (ED) up to $L = 15$ suggests a finite ground state energy (and thus a single particle gap) which is then invalidated by DMRG up to $L = 108$. The simulations were performed with the ALPS libraries [3, 4].

Note that we did *not* disprove the existence of potentially lower lying bound states.

This is, in general, a highly nontrivial task. For instance, it has been shown rigorously that the gap above the zero-energy ground state for the isotropic Heisenberg chain is given by $1 - \cos(\pi/L) = 2\sin^2 \pi/2L$ [5], which indeed agrees with the single particle excitations for $k = \frac{\pi}{L}$ given above (the additional factor of 2 follows from the definition of our Hamiltonian in terms of spinless fermions). From a rigorous point of view, this statement can only be transferred to $H^a + H^b$ which describes two non-interacting isotropic Heisenberg chains.

However, our results suggest that the additional interaction $H^{ab}$ cannot decrease the first excitation energy given by the $k = \pi/L$ magnons (which would certainly be true for weak perturbations $\varepsilon H^{ab}$, $\epsilon \ll 1$). This statement is underpinned by DMRG simulations for half filling up to $L = 108$ with very high accuracy, see Fig. 2 (a). We conclude that for OBC the energy gap closes as

$$\Delta E_0(L) \sim \frac{\pi^2}{L^2} \tag{30}$$

in all sectors $(N, \alpha)$.

We proceed with the case of periodic boundary conditions. As explained in the main text, there is a difference between the odd-odd sector $((-1)^N = +1$ and $\alpha = -1)$ with an exact zero-energy state and the remaining even-even and even-odd/odd-even sectors with finite energy ground states. Consider the odd-odd sector first. There the fermionic statistics has no effect and the energy expectation value has the form of Eq. (9) for OBC. Performing a Jordan-Wigner transformation yields the isotropic Heisenberg chain with (untwisted) PBC. Due to the survival of the SU(2)-symmetry, all statements about the ground state and the single-particle excitations carry over from the previous case of OBC; in particular the gap closing, now with $\Delta E_0(L) \sim 4\pi^2/L^2$ since the lowest-energy magnon has momentum $k = 2\pi/L$ (cf. $k = \pi/L$ for OBC). These statements can again be verified to high precision with DMRG simulations, see Fig. 2 (a).

We conclude with the sectors without zero-energy ground state, i.e., even-even and even-odd/odd-even. Here we aim at the scaling of the ground state energy $E_0(L) \propto L^{-1}$ compared to the zero-energy ground states in the odd-odd sector. In contrast to all previous cases, here we have neither access to exact ground nor excited states which complicates the analysis considerably.

We start with the even-even sector. Here one can show

$$\frac{C_1}{L} \geq E_0(L) \geq \frac{C_2}{L^2} \tag{31}$$

rigorously. The upper bound follows with the ansatz wave function $(0 \leq K \leq L)$

$$|G_0\rangle = \exp\left[i\frac{\pi}{L}\sum_{s=1}^{L} s(a_s^\dagger a_s + b_s^\dagger b_s)\right]|2K, +1\rangle \tag{32}$$

which imprints a single-particle phase field onto the equal-weight superposition and thereby satisfies the twisted boundary conditions with zero-energy in the thermodynamic limit. It can be shown by straightforward calculations that

$$\langle G_0|\,H\,|G_0\rangle = 2 \cdot 4\rho(1-\rho) \cdot L \cdot \sin^2 \frac{\pi}{2L}$$
$$\sim 2\rho(1-\rho)\pi^2 \cdot L^{-1} \tag{33}$$

with the filling $\rho = N/2L = K/L$. The energy is due to intra-chain interactions alone, the inter-chain contributions due to $H^{ab}$ vanish.

Finding the lower bound (thereby establishing the algebraic decay) is much more subtle as it makes statements about *all* possible states — in contrast to the upper bound which was derived by a single ansatz wave function. The basic line of thought is to consider the energy expectation value in Eq. (9) for a generic state

$$\langle\Psi|\,H\,|\Psi\rangle \propto \sum_{(ij)\in G} |\Psi_i - \alpha_{ij}\Psi_j|^2 \qquad (34)$$

where now $\alpha_{ij} = \pm 1$ due to the fermionic statistics and $i,j$ are indices encoding the fermion configurations $(\boldsymbol{n},\boldsymbol{m})$. $G$ can be thought of as a graph describing the possible local transformations $\Psi_i \mapsto \Psi_j$ due to the Hamiltonian $H$. One can now construct a set $\mathcal{C}$ of length-$L$ cycles $C$ in $G$ such that (i) every configuration $i$ is visited by at least one of the cycles and (ii) every cycle contains one edge $(ij)$ with $\alpha_{ij} = -1$, preventing zero-energy on each cycle. Physically, these cycles can be interpreted as virtual tunneling of a single fermion around one leg of the ladder:

$$\langle\Psi|\,H\,|\Psi\rangle \geq \sum_{C\in\mathcal{C}} \sum_{(ij)\in C} |\Psi_i - \alpha_{ij}\Psi_j|^2 \qquad (35)$$

An application of Hölder's inequality followed by the triangle inequality yields

$$\langle\Psi|\,H\,|\Psi\rangle \geq \tfrac{1}{L^2} \sum_{C\in\mathcal{C}} \sum_{i\in C} |\Psi_i|^2 \qquad (36)$$

where the single $\alpha_{ij} = -1$ is crucial. Due to the construction of $\mathcal{C}$ based on *cluster*-configurations of fermions, the double sum can be recast as sum over all configurations $i$ weighted by their number of fermion clusters $N_c(i)$,

$$\langle\Psi|\,H\,|\Psi\rangle \geq \tfrac{1}{L^2} \sum_i N_c(i)|\Psi_i|^2 = \tfrac{\langle N_c\rangle}{L^2} \geq \tfrac{1}{L^2}. \qquad (37)$$

In the last step we bluntly bounded the ground state expectation of clusters by one from below which establishes the lower bound.

The stronger statement $E_0(L) \propto L^{-1}$ is again supported by numerical simulations and analytic results known for the Heisenberg model with twisted boundary conditions [6–8]. We stress that numerics suggests in the ground state $\langle N_c\rangle \sim L$ providing the numerically observed lower bound $L^{-1}$. Note that a bounded cluster number, $\langle N_c\rangle \leq$ const, would indicate phase separation.

Finally, consider the closed ladder with odd total filling, i.e., the even-odd/odd-even sectors. We point out that the previous sketch providing the lower bound $L^{-2}$ only relies on the existence of a single chain with even filling. It is therefore still applicable in the present case and restricts any decay of the ground state energy to be algebraic. However, the former ansatz wave function (32) yields a finite energy expectation as only one of the two chains requires a single particle phase due to the even filling. The naïve ansatz $|G_0\rangle = \exp\left[i\tfrac{\pi}{L}\sum_{s=1}^L s a_s^\dagger a_s\right]|2K+1,+1\rangle$ (here for $\alpha=+1$) with a single-particle phase only in the upper chain fails with an extensively growing energy expectation due to the pair-phase locking of the interaction $H^{ab}$.

In this picture of single particle-phases (naïvely supported by bosonization) one might therefore conclude that the ground state energy is finite, $\lim_{L\to\infty} E_0(L) > 0$, which then established a single particle gap, i.e. a finite difference of ground state energies between even and odd total filling. This conclusion is even supported by exact diagonalization for $\rho = 0.5$ up to $L = 15$ where the ground state energy seems to approach a finite gap. However, DMRG up to $L = 108$ proves this conclusion wrong and suggests clearly a vanishing of $E_0(L)$ with $L^{-1}$ for any fixed filling, see Fig. 2 (b). The non-existence of a single-particle gap is presumably owed to the "double" criticality of the very peculiar point $H$ under consideration, in that it is not only a gapless phase but also on the brink of phase separation.

### Wire Networks

For the braiding procedure described in the main text (Fig. 5) a much more complicated setup than the "simple" two-leg ladder was used. Here we comment on this in more detail: We show that a much wider class of Hamiltonians, defined on "wire networks", can be described in exactly the same way as the simple two-leg ladder; including the degenerate zero-energy ground states given as equal-weight superpositions and the simple evaluation of correlators/observables in terms of PsBCs.

Examples of possible network topologies accessible by these techniques are shown in Fig. 3. In (a) two wires coupled by a common closed bath chain give rise to a two-dimensional ground state degeneracy, characterized by the parities of the two subchains. In (b) a more complex patch from a stacked wire setup is shown. The more intricate setup in (c) will be used in the following to illustrate the construction of a generic wire Hamiltonian derived from the two-leg version of the main text.

In fact, for any network of $g$ open and $g'$ closed (single) chains, sectionally connected as double wires, one can write down a generalized version of the wire Hamiltonian $H$ such that in any filling sector $N$, there are exactly $2^{g-1}$ ($g\geq 1$) zero-energy states $|N;\alpha_1,\ldots,\alpha_{g-1}\rangle$ given as the equal-weight superpositions of fermion configurations with the $i$th subchain parity fixed at $\alpha_i = \pm 1$. In the special case of $g = 1$, the ground state $|N\rangle$ for each filling $N$ is uniquely determined. The parity of *closed* single chains is fixed at $\alpha = -1$ to make the ground state energy vanish identically, so only *open* chains contribute to the ground state degeneracy. Note that $g = E/2$ where $E$ denotes the number of open subchain ends, as used in the main text.

For instance, consider the $g = 2$ wire network in Fig. 3 (a) where the upper "bath" chain is closed and therefore has fixed parity $\alpha_0 = -1$. Then the two open

chain parities $\alpha_1$ and $\alpha_2$ are constrained by the total parity via $\alpha_1 \alpha_2 = -(-1)^N$. Hence the ground state space is two-fold degenerate for fixed filling $N$, as illustrated by the two possible subparity sectors. Although the degeneracy is the same as for the simple ladder setup used in the main text, here the edge states are spatially separated and therefore resilient even against TRB perturbations.

To write down the local Hamiltonian featuring the properties sketched above, we start with an arbitrary wire network $\mathcal{L}$ of fermionic sites (for the sake of clarity, think of the example in Fig. 3 (c)). Formally, $\mathcal{L} = (\mathbb{V}, \mathcal{C}, \mathcal{P})$ is defined on a (so far unordered) set $\mathbb{V}$ of vertices $s$ (fermionic sites) as a collection $\mathcal{C}$ of unoriented chains $C$ (open or closed) and a collection $\mathcal{P}$ of pairing sections $P$ describing the sectional attachment of two chains $C_1$ and $C_2$ to form a two-leg ladder in some segment.

Note that a single chain $C$ can be oriented in two different ways while preserving the locality of the sites; this defines a *direction* of each single chain (which is a completely unphysical, that is, gauge choice). In Fig. 3 (c) a possible choice is given by the black arrows. Note that the only relevant manifestation of this (gauge) directions is defined on pairing segments $P \in \mathcal{P}$ as a binary function $\chi : \mathcal{P} \to \{-1, +1\}$ via $\chi(P) = +1$ $(-1)$ if the orientations of the two bonded chains are parallel (antiparallel). In the following, (abstract) wire networks $\mathcal{L}$ are always thought of being augmented by an orientation function $\chi$.

After these formal preliminaries, we can write down the Hamiltonian that governs the physics on the wire network $\mathcal{L}$ as follows:

$$H[\mathcal{L}] = \sum_{C \in \mathcal{C}} H'[C] + \sum_{P \in \mathcal{P}} H''[P] \qquad (38)$$

where $H'$ describes the intra-chain physics,

$$H'[C] = \sum_{e \in C} A_e (\mathbb{1} + A_e) \qquad (39)$$

with the single particle hopping $A_e = a_s a_p^\dagger + a_p a_s^\dagger$ on edge $e = (s, p)$, and $H''$ is responsible for the inter-chain interaction,

$$H''[P] = \sum_{f \in P} B_f (\chi(P) + B_f) \qquad (40)$$

with the known pair hopping between the two chains $B_f = a_s^\dagger a_p^\dagger a_r a_q + a_r^\dagger a_q^\dagger a_s a_p$. Here $f = (s, p; q, r)$ defines a "face" on the ladder segment with corners $s, p$ on the "upper" and $q, r$ on the "lower" chain, Fig. 3 (c). Note that the choice of chain orientations affects only the definition of $H''$ which is rooted in the invariance of $A_e$ with respect to orientation inversion on a single chain ($s \leftrightarrow p$), in contrast to the variance $B_f \leftrightarrow -B_f$ induced by orientation inversion on one of the two chains (e.g., $s \leftrightarrow p$).

We point out that in the framework of the fermion algebra $\{a_s\}_{s \in \mathbb{V}}$, the gauge transformation $a_s \to i a_s$ for $s \in C$ inverts all signs of $\chi(P)$ for pairing sections $P$ including chain $C$. In this sense, the orientation arrows of

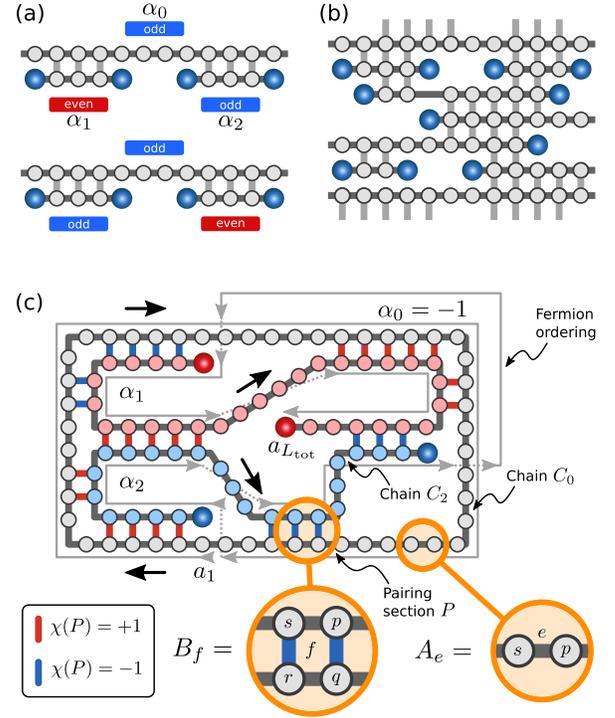

Figure 3. *Wire Networks.* (a) Simple wire network with two open chains connected by a common closed "bath" chain fixed at odd parity. The ground states are characterized by the subchain parities. (b) Patch of a more complex network of stacked chains. The degeneracy grows with the number of open chain ends $E$. (c) Example of a more intricate ladder setup with a closed "bath" chain fixed at $\alpha_0 = -1$ subchain parity and the two open subchain parities $\alpha_1, \alpha_2$. There are two exact zero-energy ground states in each particle number sector characterized by $-\alpha_1 \alpha_2 = (-1)^N$. This scheme is used to introduce a generic version of the two-leg ladder Hamiltonian introduced in the main text.

single chains can be inverted by the U(1)-rotation of the fermions on the chain in question. Then it is obvious that the choice of orientations (and thereby $\chi$) has no physical significance as the resulting Hamiltonians are related by gauge transformations. It is important to stress that while the choice of *orientations* is unrestricted, the relative orientations $\chi$ are implicitly given by this choice and the topology of the wire network. That is to say, there are different gauge equivalence classes of Hamiltonians for a given wire network $\mathcal{L}$, and only the representatives $H[\mathcal{L}]$ (of a specific class) defined above give rise to degenerate zero-energy states.

For instance, there are network topologies where it is *not* possible to gauge all relative orientations to $\chi(P) = +1$ (which would be the naïve generalization of our paradigmatic two-leg ladder to networks). One such example is given in Fig. 3 (c) where the colors of inter-chain pair interactions encode the sign of $\chi(P)$ with respect to the orientations given by the black arrows. Note that there is no choice of orientations so that all ladder seg-

ments feature parallel orientations.

It remains to be shown that the ground state space of $H[\mathcal{L}]$ features degenerate zero-energy states given by equal-weighted superpositions and characterized by the parities of the open subchains in $\mathcal{L}$. To this end, a representation of the fermion algebra on the Hilbert space is required. This is achieved by defining a number state basis of the Fock space, based on a total ordering of the fermion sites. A convenient choice for the construction of the ground states is to order fermions along single chains and parallel to the previously chosen orientations. The sequence of the single chains relative to each other is not important as long as only parity conserving operations between the chains are concerned (which is true for $H[\mathcal{L}]$). Note that the sequence of chains *is* important for the evaluation of subchain parity violating observables/correlators such as inter-chain Green's functions and single particle hopping. A possible ordering is highlighted in Fig. 3 (c).

The construction of the ground states follows from exactly the same reasoning as described above for the simple two-leg ladder, generalized to multiple connected chains. The single particle hopping within the chains can still be described on the previously defined Fock basis as

$$A_e(\mathbb{1} + A_e)\ket{\boldsymbol{n}} = \ket{\boldsymbol{n}} - \ket{\varepsilon_e(\boldsymbol{n})} \tag{41}$$

with, $e = (i, i+1)$,

$$\varepsilon_e(\ldots, n_i, n_{i+1}, \ldots) = (\ldots, n_{i+1}, n_i, \ldots) \tag{42}$$

where we no longer split the fermion configuration into two subchain configurations $\boldsymbol{n}$ and $\boldsymbol{m}$ but consider the total configurations $\boldsymbol{n} \in \{0,1\}^{L_{\text{tot}}}$ with the total (single) chain length $L_{\text{tot}} = \sum_{i=1}^{g+g'} L_i$ of all $g+g'$ chains. For the two-leg ladder one has $g = 2$, $g' = 0$ and $L_{\text{tot}} = L + L = 2L$.

For the inter-chain pair interaction one finds

$$B_f(\chi(P) + B_f)\ket{\boldsymbol{n}} = \ket{\boldsymbol{n}} - \ket{\tau_f(\boldsymbol{n})} \tag{43}$$

with

$$\tau_f(\ldots, n_i, n_{i+1}, \ldots, n_j, n_{j+1}, \ldots)$$
$$= (\ldots, n_j, n_{j+1}, \ldots, n_i, n_{i+1}, \ldots) \tag{44}$$

if $n_i = n_{i+1}$ and $n_j = n_{j+1}$, and $\tau_f = \text{Id}$ otherwise. Now it becomes clear that the sign $\chi(P)$ in $H''$ compensates for the signs due to fermion pair hopping, which are different for hopping between chains of parallel and antiparallel order (the latter includes an exchange of the pair fermions). The effect of $H[\mathcal{L}]$ (depending on the chain orientations through $\chi$) on the used Fock basis $\ket{\boldsymbol{n}}$ (also depending on the chain orientations) is therefore invariant under the choice of chain orientations and has the same form as for the two-leg ladder.

In conclusion, the energy expectation still has the generic form

$$\bra{\Psi} H[\mathcal{L}] \ket{\Psi} \propto \sum_{(\boldsymbol{n},\boldsymbol{n}') \in G} |\Psi(\boldsymbol{n}) - \Psi(\boldsymbol{n}')|^2 \tag{45}$$

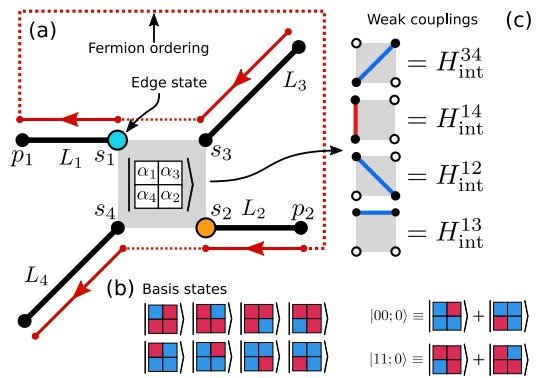

Figure 4. *Braiding scheme.* (a) Braiding setup with four chains $L_i$, $i = 1, 2, 3, 4$. The "bath" chain is omitted as it is not relevant for the low-energy dynamics. The chosen fermion ordering defining the number state basis is highlighted red. $\alpha_i$ denote the subchain parities fixed to $\alpha_1\alpha_2\alpha_3\alpha_4 = -1$. (b) 8 basis states of the ground state space in which the braiding dynamic takes place. The linear combinations $\ket{00;0}$ and $\ket{11;0}$ are exchanged by the braiding operation. (c) Weak single-particle couplings used for the braiding. Blue (red) edges denote time-reversal invariant (breaking) single-particle hopping.

where $(\boldsymbol{n}, \boldsymbol{n}')$ encodes the possibility to transform the fermion configurations $\boldsymbol{n} \leftrightarrow \boldsymbol{n}'$ into each other by an elementary process of $H[\mathcal{L}]$ (i.e., a single particle hopping within a chain or a pair hopping between chains in a bonding region). Note that Eq. (45) is only valid for number states with odd subchain parity $\alpha_C = -1$ for *closed* chains $C$.

The zero-energy ground states of the positive semidefinite Hamiltonian $H[\mathcal{L}]$ are therefore given by states with constant weights on all connected components of $G$, that is, on fermion configurations with fixed total particle number $N$ and fixed subchain parities $\alpha_i$, $i = 1, \ldots, g-1$. This gives rise to the claimed $2^{g-1}$ fold degeneracy in each filling sector $N$, spanned by the ground states $\ket{N; \alpha_1, \ldots, \alpha_{g-1}}$. It is now easy to see that arbitrary correlators/observables still can be expressed in terms of PsBCs; the splitting degree of which may be much larger though.

A detailed analysis of such wire networks (e.g., effects due to perturbations of inter-chain single-particle hopping) might be a promising field for future work.

## Braiding

The setup used in the main text, see Fig. 5 (a), is an example of a wire network, the ground states of which are given by the equal-weight superpositions of all fermion configurations with total number $N$ and subchain parities $\alpha_i$, $i = 1, 2, 3, 4$. The parity of the "bath" chain connecting the four open chains can be fixed so that $\alpha_1\alpha_2\alpha_3\alpha_4 = -1$ or $\alpha_0 = -(-1)^N$, and is of no interest in

the following. Due to the previous constraint on the subchain parities, the ground state space can be restricted to the eight states $\left|\begin{smallmatrix}\alpha_1\alpha_3\\\alpha_4\alpha_2\end{smallmatrix}\right\rangle$ with fixed particle number $N$ and bath parity $\alpha_0$, see Fig. 4 (b).

For the braiding we consider the time-dependent Hamiltonian

where the time-independent part $H_0$ defines the wire network without couplings between the endpoints of subchains $L_i$ (including the bath chain $L_0$ absent in Fig. 4 (a)) and the weak single-particle coupling $H_{\mathrm{int}}(t)$ describes the time-dependent braiding procedure within the above described ground state sector of $H_0$ (Fig. 4 (c)).

$$H(t) = H_0 + \varepsilon/L^2\, H_{\mathrm{int}}(t) \qquad (46)$$

The Hamiltonian $H_{\mathrm{int}}(t)$ couples the four chains $L_i$, $i=1,2,3,4$, via single-particle hopping as follows:

$$\tilde{H}_{\mathrm{int}}(t) = \begin{cases} (1-\tau)\,H_{\mathrm{int}}^{34} + \tau\,H_{\mathrm{int}}^{14} & \text{for } 0\leq t<1 \text{ and } \tau = t-0 \\ (1-\tau)\,H_{\mathrm{int}}^{14} + \tau\,H_{\mathrm{int}}^{12} & \text{for } 1\leq t<2 \text{ and } \tau = t-1 \\ (1-\tau)\,H_{\mathrm{int}}^{12} + \tau\,H_{\mathrm{int}}^{13} & \text{for } 2\leq t<3 \text{ and } \tau = t-2 \\ (1-\tau)\,H_{\mathrm{int}}^{13} + \tau\,H_{\mathrm{int}}^{14} & \text{for } 3\leq t<4 \text{ and } \tau = t-3 \end{cases} \qquad (47)$$

and we set $H_{\mathrm{int}}(t) \equiv \tilde{H}_{\mathrm{int}}(t \bmod 4)$ to allow for a double exchange (that is, full braiding) of the two edge states located on the inner endpoints $s_1$ and $s_2$ of chains $L_1$ and $L_2$ for $0 \leq t \leq 8$. The chain coupling Hamiltonians are defined analogous to the intra-chain coupling $A_i(\mathbb{1}+A_i)$, namely

$$H_{\mathrm{int}}^{34} = a_{s_3}a_{s_4}^\dagger + a_{s_4}a_{s_3}^\dagger + n_{s_3}(\mathbb{1}-n_{s_4}) + n_{s_4}(\mathbb{1}-n_{s_3}) \qquad (48\mathrm{a})$$

$$H_{\mathrm{int}}^{14} = i a_{s_1}a_{s_4}^\dagger - i a_{s_4}a_{s_1}^\dagger + n_{s_1}(\mathbb{1}-n_{s_4}) + n_{s_4}(\mathbb{1}-n_{s_1}) \qquad (48\mathrm{b})$$

$$H_{\mathrm{int}}^{12} = a_{s_1}a_{s_2}^\dagger + a_{s_2}a_{s_1}^\dagger + n_{s_1}(\mathbb{1}-n_{s_2}) + n_{s_2}(\mathbb{1}-n_{s_1}) \qquad (48\mathrm{c})$$

$$H_{\mathrm{int}}^{13} = a_{s_1}a_{s_3}^\dagger + a_{s_3}a_{s_1}^\dagger + n_{s_1}(\mathbb{1}-n_{s_3}) + n_{s_3}(\mathbb{1}-n_{s_1}) \qquad (48\mathrm{d})$$

where $s_i$ denotes the terminating fermion sites on chain $L_i$, see Fig. 4 (a). Note that the single-particle hopping in $H_{\mathrm{int}}^{14}$ breaks time-reversal symmetry — which is necessary to create a finite overlap between ground states in the thermodynamic limit for couplings between chains with reversed fermion ordering.

In the ground state basis, the coupling Hamiltonian matrix elements read as follows (here exemplarily for $H_{\mathrm{int}}^{34}$)

$$\left\langle\begin{smallmatrix}\beta_1\beta_3\\\beta_4\beta_2\end{smallmatrix}\right| H_{\mathrm{int}}^{34} \left|\begin{smallmatrix}\alpha_1\alpha_3\\\alpha_4\alpha_2\end{smallmatrix}\right\rangle = \left[\delta_{\alpha,\beta}\mathcal{N}_\alpha^{-1} + \delta_{\alpha_1,\beta_1}\delta_{\alpha_2,\beta_2}\delta_{\alpha_3,-\beta_3}\delta_{\alpha_4,-\beta_4}\mathcal{N}_\alpha^{-1/2}\mathcal{N}_\beta^{-1/2}(-\alpha_1\alpha_2)\right] \qquad (49)$$

$$\times \left[ \begin{pmatrix} L_1, L_2, L_3-1, L_4-1, L_0 \\ \alpha_1, \alpha_2, \alpha_3, -\alpha_4 \end{pmatrix}_{N-1} + \begin{pmatrix} L_1, L_2, L_3-1, L_4-1, L_0 \\ \alpha_1, \alpha_2, -\alpha_3, \alpha_4 \end{pmatrix}_{N-1} \right] \qquad (50)$$

with the shorthand notation $\delta_{\alpha,\beta} \equiv \prod_{i=1}^4 \delta_{\alpha_i,\beta_i}$ and the normalizing factor

$$\mathcal{N}_\alpha = \begin{pmatrix} L_1, L_2, L_3, L_4, L_0 \\ \alpha_1, \alpha_2, \alpha_3, \alpha_4 \end{pmatrix}_N. \qquad (51)$$

Using that

$$\begin{pmatrix} L_1, L_2, L_3, L_4, L_0 \\ \alpha_1, \alpha_2, \alpha_3, \alpha_4 \end{pmatrix}_N^{-1} \begin{pmatrix} L_1, L_2, L_3-1, L_4-1, L_0 \\ \beta_1, \beta_2, \beta_3, \beta_4 \end{pmatrix}_{N-1} \longrightarrow \rho(1-\rho) \qquad (52)$$

for $L_i \to \infty$ up to exponentially small $\alpha,\beta$-dependent terms, leads to the expressions

$$\left\langle\begin{smallmatrix}\beta_1\beta_3\\\beta_4\beta_2\end{smallmatrix}\right| H_{\mathrm{int}}^{34} \left|\begin{smallmatrix}\alpha_1\alpha_3\\\alpha_4\alpha_2\end{smallmatrix}\right\rangle \sim \delta_{\alpha,\beta}\cdot 1 + \delta_{\alpha_1,\beta_1}\delta_{\alpha_2,\beta_2}\delta_{\alpha_3,-\beta_3}\delta_{\alpha_4,-\beta_4} \cdot (-\alpha_1\alpha_2) \qquad (53\mathrm{a})$$

$$\left\langle\begin{smallmatrix}\beta_1\beta_3\\\beta_4\beta_2\end{smallmatrix}\right| H_{\mathrm{int}}^{14} \left|\begin{smallmatrix}\alpha_1\alpha_3\\\alpha_4\alpha_2\end{smallmatrix}\right\rangle \sim \delta_{\alpha,\beta}\cdot 1 + \delta_{\alpha_1,-\beta_1}\delta_{\alpha_2,\beta_2}\delta_{\alpha_3,\beta_3}\delta_{\alpha_4,-\beta_4} \cdot (-i\alpha_1\alpha_2) \qquad (53\mathrm{b})$$

$$\left\langle\begin{smallmatrix}\beta_1\beta_3\\\beta_4\beta_2\end{smallmatrix}\right| H_{\mathrm{int}}^{12} \left|\begin{smallmatrix}\alpha_1\alpha_3\\\alpha_4\alpha_2\end{smallmatrix}\right\rangle \sim \delta_{\alpha,\beta}\cdot 1 + \delta_{\alpha_1,-\beta_1}\delta_{\alpha_2,-\beta_2}\delta_{\alpha_3,\beta_3}\delta_{\alpha_4,\beta_4} \cdot (+\alpha_1\alpha_2) \qquad (53\mathrm{c})$$

$$\left\langle\begin{smallmatrix}\beta_1\beta_3\\\beta_4\beta_2\end{smallmatrix}\right| H_{\mathrm{int}}^{13} \left|\begin{smallmatrix}\alpha_1\alpha_3\\\alpha_4\alpha_2\end{smallmatrix}\right\rangle \sim \delta_{\alpha,\beta}\cdot 1 + \delta_{\alpha_1,-\beta_1}\delta_{\alpha_2,\beta_2}\delta_{\alpha_3,-\beta_3}\delta_{\alpha_4,\beta_4} \cdot (-1). \qquad (53\mathrm{d})$$

These results are valid for $L_i \to \infty$ for all chains $i$ and fixed filling $\rho$. We omit the global factor of $2\rho(1-\rho)$ which is justified as it is independent of the parities $\alpha_i$ up to exponentially small corrections.

We integrate the propagator

$$U(t) = \exp\left[-i\int_0^t \mathrm{d}\tau\, H_{\mathrm{int}}(\tau)\right] \qquad (54)$$

numerically and thereby end up with a non-trivial holon-

omy $U_{s_1s_2} \equiv U(4)$ on the four-fold degenerate ground state space describing the action of exchanging ($U_{s_1s_2}$) and braiding ($U_{s_1s_2}^2$) the two localized edge states on $s_1$ and $s_2$.

The ground state space at $t = 0 \mod 4$ is characterized by a coupling of chains $L_3$ and $L_4$ and therefore spanned by the four states $|\alpha_1, \alpha_2\rangle |\alpha_3\alpha_4\rangle_+$ with $\alpha_1\alpha_2\alpha_3\alpha_4 = -1$ defined as

$$|\alpha_1, \alpha_2\rangle |\alpha_3\alpha_4\rangle_+ \equiv \begin{vmatrix} \alpha_1\alpha_3 \\ \alpha_4\alpha_2 \end{vmatrix} + \begin{vmatrix} \alpha_1 & -\alpha_3 \\ -\alpha_4 & \alpha_2 \end{vmatrix} \quad (55)$$

up to normalizations. The action of the braiding $U_{s_1s_2}^2$ is then diagonal in the $\alpha_3\alpha_4$ blocks and reads

$$U_{s_1s_2}^2 |\alpha_1, \alpha_2\rangle |\alpha_3\alpha_4\rangle_+ = |-\alpha_1, -\alpha_2\rangle |\alpha_3\alpha_4\rangle_+ \quad (56)$$

up to additional phases.

In order to understand this result more thoroughly, it is useful to cast these expressions into the language of the original (quadratic) Majorana chain, keeping in mind that due to the interactions there are no "fermion modes" or "Majorana modes" describing the ground states — this is just semantic sugar.

Remember that the delocalized, fermionic edge mode of the original Majorana chain is empty (occupied) if the total chain parity is odd (even). In this linguistic framework, we may call the edge state of a subchain in our interacting model empty (occupied) if its subchain parity is odd (even). That is, the occupation $n_1$ of the "fermion mode" defined by the pair of "Majorana modes" at $p_1$ and $s_1$, see Fig. 4 (a), corresponds to the subchain parity $\alpha_1$ via $\alpha_1 = -(-1)^{n_1}$; and the same for $L_2$. This defines the new naming scheme $|n_1, n_2\rangle |n_{34}\rangle_+ = |\alpha_1, \alpha_2\rangle |\alpha_3\alpha_4\rangle_+$ with $n_{1,2} = 0, 1$ and $\alpha_3\alpha_4 = -(-1)^{n_{34}}$. Then the action of braiding the "Majoranas" $s_1$ and $s_2$ reads

$$U_{s_1s_2}^2 |0, 0\rangle |0\rangle_+ = |1, 1\rangle |0\rangle_+ \quad (57)$$

which would be interpreted in the quadratic mean-field theory of the Majorana chain as creation of a fermion pair out of the superconducting condensate by braiding two Majorana modes.

Pairing the Majorana modes of two *different* Majorana chains into a fermion mode, one finds that the symmetric (antisymmetric) linear combination of the two parity subsectors admissible by the fixed total parity corresponds to an empty (occupied) mode. Translated into our case this means that the occupation $m$ of the "fermion mode" created from "Majorana modes" at $s_1$ and $s_2$ is given via

$$|m\rangle |m', n_{34}\rangle \quad (58)$$
$$\equiv [|\alpha_1, \alpha_2\rangle + (-1)^m |-\alpha_1, -\alpha_2\rangle] |\alpha_3\alpha_4\rangle_+$$

which, of course, is just a basis transformation. Here $\alpha_1\alpha_2 = -(-1)^{m'}$ describes the occupation $m'$ of the "fermion mode" defined by $p_1$ and $p_2$. In this basis, the action of the braiding is particularly simple, that is, diagonal

$$U_{s_1s_2}^2 |m\rangle |m', n_{34}\rangle = (-1)^m |m\rangle |m', n_{34}\rangle \quad (59)$$

up to a phase independent of $m$. This is exactly the relative phase one would expect from rotating a fermion mode by $2\pi$ with occupation $m$.

---

* [nicolai@itp3.uni-stuttgart.de](nicolai@itp3.uni-stuttgart.de)
[1] A. Kitaev, AIP Conference Proceedings **1134**, 22 (2009).
[2] S. Ryu, A. P. Schnyder, A. Furusaki, and A. W. W. Ludwig, New Journal of Physics **12**, 065010 (2010).
[3] M. Dolfi, B. Bauer, S. Keller, A. Kosenkov, T. Ewart, A. Kantian, T. Giamarchi, and M. Troyer, Computer Physics Communications **185**, 3430 (2014).
[4] B. Bauer, L. D. Carr, H. G. Evertz, A. Feiguin, J. Freire, S. Fuchs, L. Gamper, J. Gukelberger, E. Gull, S. Guertler, A. Hehn, R. Igarashi, S. V. Isakov, D. Koop, P. N. Ma, P. Mates, H. Matsuo, O. Parcollet, G. Pawłowski, J. D. Picon, L. Pollet, E. Santos, V. W. Scarola, U. Schollwöck, C. Silva, B. Surer, S. Todo, S. Trebst, M. Troyer, M. L. Wall, P. Werner, and S. Wessel, Journal of Statistical Mechanics: Theory and Experiment **2011**, P05001 (2011).
[5] T. Koma and B. Nachtergaele, Letters in Mathematical Physics **40**, 1 (1997).
[6] B. S. Shastry and B. Sutherland, Phys. Rev. Lett. **65**, 243 (1990).
[7] B. Sutherland and B. S. Shastry, Phys. Rev. Lett. **65**, 1833 (1990).
[8] C. J. Hamer, G. R. W. Quispel, and M. T. Batchelor, Journal of Physics A: Mathematical and General **20**, 5677 (1987).



\end{document}